\documentclass[reprint, superscriptaddress, floatfix, prb]{revtex4-2}

\usepackage{geometry}
\makeatletter
\def\@hangfrom@section#1#2#3{\@hangfrom{#1#2}#3}
\def\@hangfroms@section#1#2{#1#2}
\makeatother

\usepackage{siunitx}
\usepackage[english]{babel}
\usepackage[utf8]{inputenc}

\usepackage{float}
\usepackage{wrapfig}
\usepackage{graphicx}   
\graphicspath{{/figures}}

\usepackage{amsmath}
\usepackage{amssymb}
\usepackage{empheq}
\usepackage{braket}
\usepackage{bbm}
\usepackage{bm}
\usepackage{textcomp}
\usepackage{comment}

\usepackage[section]{placeins}
\usepackage{xcolor}
\usepackage{layouts}

\usepackage[colorlinks, hypertexnames=false]{hyperref}
\definecolor{navygray}{RGB}{110,140,170}
\hypersetup{
citecolor={navygray},
linkcolor={navygray},
urlcolor={navygray},
}
\usepackage[all]{hypcap}

\DeclareRobustCommand{\SkipTocEntry}[5]{}

\newcommand{\eref}[1]{\hyperref[#1]{{Eq.~\ref{#1}}}}  
\newcommand{\eqsref}[1]{\hyperref[#1]{{Eqs.~\ref{#1}}}}  
\newcommand{\fref}[1]{\hyperref[#1]{{Fig.~\ref{#1}}}}
\newcommand{\frefadd}[2]{\hyperref[#1]{{Fig.~\ref*{#1}#2}}}

\begin{document}

\title{Two-level system hyperpolarization using a quantum Szilard engine}

\author{Martin Spiecker}
\email{martin.spiecker@kit.edu}
\affiliation{PHI,~Karlsruhe~Institute~of~Technology,~76131~Karlsruhe,~Germany}
\affiliation{IQMT,~Karlsruhe~Institute~of~Technology,~76344~Eggenstein-Leopoldshafen,~Germany}

\author{Patrick Paluch}
\affiliation{PHI,~Karlsruhe~Institute~of~Technology,~76131~Karlsruhe,~Germany}
\affiliation{IQMT,~Karlsruhe~Institute~of~Technology,~76344~Eggenstein-Leopoldshafen,~Germany}

\author{Nicolas Gosling}
\affiliation{IQMT,~Karlsruhe~Institute~of~Technology,~76344~Eggenstein-Leopoldshafen,~Germany}

\author{Niv Drucker}
\affiliation{Quantum Machines, 67443 Tel Aviv-Yafo, Israel}

\author{Shlomi Matityahu}
\affiliation{TKM,~Karlsruhe~Institute~of~Technology,~76131~Karlsruhe,~Germany}

\author{Daria Gusenkova}
\affiliation{PHI,~Karlsruhe~Institute~of~Technology,~76131~Karlsruhe,~Germany}
\affiliation{IQMT,~Karlsruhe~Institute~of~Technology,~76344~Eggenstein-Leopoldshafen,~Germany}

\author{Simon Günzler}
\affiliation{PHI,~Karlsruhe~Institute~of~Technology,~76131~Karlsruhe,~Germany}
\affiliation{IQMT,~Karlsruhe~Institute~of~Technology,~76344~Eggenstein-Leopoldshafen,~Germany}

\author{Dennis Rieger}
\affiliation{PHI,~Karlsruhe~Institute~of~Technology,~76131~Karlsruhe,~Germany}
\affiliation{IQMT,~Karlsruhe~Institute~of~Technology,~76344~Eggenstein-Leopoldshafen,~Germany}

\author{Ivan Takmakov}
\affiliation{PHI,~Karlsruhe~Institute~of~Technology,~76131~Karlsruhe,~Germany}
\affiliation{IQMT,~Karlsruhe~Institute~of~Technology,~76344~Eggenstein-Leopoldshafen,~Germany}

\author{Francesco Valenti}
\affiliation{IQMT,~Karlsruhe~Institute~of~Technology,~76344~Eggenstein-Leopoldshafen,~Germany}

\author{Patrick Winkel}
\affiliation{PHI,~Karlsruhe~Institute~of~Technology,~76131~Karlsruhe,~Germany}
\affiliation{IQMT,~Karlsruhe~Institute~of~Technology,~76344~Eggenstein-Leopoldshafen,~Germany}

\author{Richard Gebauer}
\affiliation{IPE,~Karlsruhe~Institute~of~Technology,~76344~Eggenstein-Leopoldshafen,~Germany}

\author{Oliver Sander}
\affiliation{IPE,~Karlsruhe~Institute~of~Technology,~76344~Eggenstein-Leopoldshafen,~Germany}

\author{Gianluigi Catelani}
\affiliation{ \mbox{JARA,~Forschungszentrum J{\"u}lich, 52425 J{\"u}lich, Germany}}

\author{Alexander Shnirman}
\affiliation{IQMT,~Karlsruhe~Institute~of~Technology,~76344~Eggenstein-Leopoldshafen,~Germany}
\affiliation{TKM,~Karlsruhe~Institute~of~Technology,~76131~Karlsruhe,~Germany}

\author{Alexey V. Ustinov}
\affiliation{PHI,~Karlsruhe~Institute~of~Technology,~76131~Karlsruhe,~Germany}
\affiliation{IQMT,~Karlsruhe~Institute~of~Technology,~76344~Eggenstein-Leopoldshafen,~Germany}

\author{Wolfgang Wernsdorfer}
\affiliation{PHI,~Karlsruhe~Institute~of~Technology,~76131~Karlsruhe,~Germany}
\affiliation{IQMT,~Karlsruhe~Institute~of~Technology,~76344~Eggenstein-Leopoldshafen,~Germany}

\author{Yonatan Cohen}
\affiliation{Quantum Machines, 67443 Tel Aviv-Yafo, Israel}

\author{Ioan M. Pop}
\email{ioan.pop@kit.edu}
\affiliation{PHI,~Karlsruhe~Institute~of~Technology,~76131~Karlsruhe,~Germany}
\affiliation{IQMT,~Karlsruhe~Institute~of~Technology,~76344~Eggenstein-Leopoldshafen,~Germany}

\date{\today}

\begin{abstract}
	The innate complexity of solid-state physics exposes superconducting quantum circuits to interactions with uncontrolled degrees of freedom degrading their coherence. By implementing a quantum Szilard engine with an active feedback control loop, we show that a superconducting fluxonium qubit is coupled to a two-level system (TLS) environment of unknown origin, with a relatively long intrinsic energy relaxation time exceeding \SI{50}{ms}. The
	TLSs can be cooled down, resulting in a four times lower qubit population, or they can be heated to manifest themselves as a negative-temperature environment corresponding to a qubit population of $\sim\SI{80}{\percent}$. We show that the TLSs and qubit are the dominant loss mechanism for each other and that qubit relaxation is independent of the TLS populations. Understanding and mitigating TLS environments is, therefore, not only crucial to improve the qubit lifetimes but also to avoid non-Markovian qubit dynamics.
\end{abstract}

\maketitle

Although tremendous progress has been made to improve the coherence of superconducting qubits, they still have to cope with various loss and decoherence mechanisms, certainly to the chagrin of quantum computing scientists, but also to the joy of mesoscopic physicists. 
The relentless interactions between superconducting hardware and its environment motivate the development of quantum error correction using stabilizer codes on the one hand~\cite{Ofek2016Aug, Vuillot2019Mar, GoogleQ2021Jul, Cai2021Jan}, and deepen our understanding of mesoscopic processes on the other hand~\cite{Grabovskij2012Oct, Barends2013, Riste2013May, Pop2014Apr, Gustavsson2016Dec, Gruenhaupt2018Sep, Serniak2018Oct, Chu2018Nov, Graaf2020, Wilen2021Jun, Glazman2021}. 
In the past, numerous strategies have been conceived to study and mitigate decoherence from various sources, from defects in dielectrics to non-thermal excitations~\cite{Siddiqi2021Oct}. 
A major source can be attributed to the wide class of TLSs in the qubit environment. Weakly coupled TLSs may be investigated by saturation pulses~\cite{Kirsh2017Jun, Andersson2021Jan} while strongly coupled TLSs may even be operated coherently via the superconducting qubit~\cite{Wang2013Apr, Lisenfeld2019}. Moreover, it has been shown that a sequence of repeated $\pi$-pulses can change the superconducting qubit's environment, which was interpreted as diffusion of superconducting quasiparticles away from the qubit's junctions~\cite{Gustavsson2016Dec}. 

\begin{figure*}[ht!]
	\begin{center}
		\includegraphics[scale=1.0]{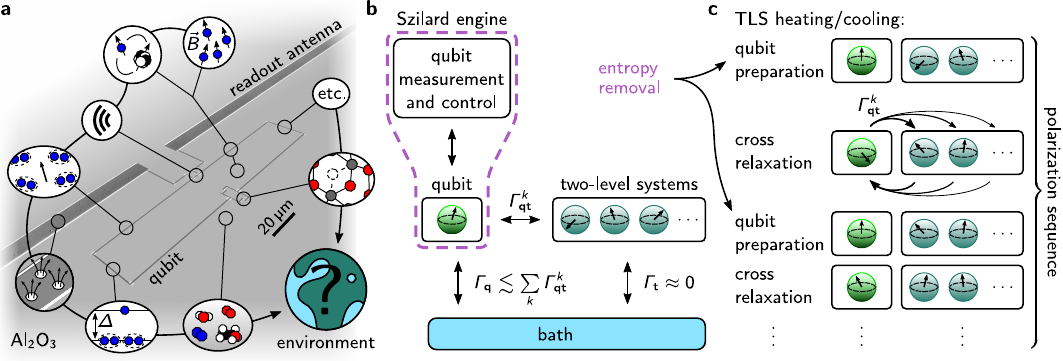}
		\caption{\textbf{The superconducting qubit, its environment and the working principle of the Szilard engine.}
			\textbf{a} Schematic drawing of the fluxonium qubit inductively coupled to its readout antenna. The rich environment typical for superconducting circuits is shown schematically and includes (counterclockwise): free electronic spins that may be Zeeman split by an external magnetic field \cite{Samkharadze2016, Borisov2020} or via the hyperfine interaction \cite{Quintana2017}, radiation loss into the readout and qubit drive ports \cite{Reed2010May} or into spurious modes including phonons \cite{vonLupke2022}, Shiba spins \cite{Yang2020Sep}, trapped vortices \cite{Vool2014, Nsanzineza2014Sep}, quasiparticles \cite{Glazman2021Jun}, absorbed molecules on the surface\cite{Kumar2016Oct}, dielectric TLSs \cite{Muller2019Oct}. The fluxonium is implemented with granular aluminum and a SQUID (superconducting quantum interference device) junction~\cite{Gruenhaupt2019, Gusenkova2021}. \textbf{b} 
			The qubit environment can be modeled as a collection of polarizable TLSs and a global bath responsible for the so-called “intrinsic loss” of both qubit ($\Gamma_\text{q}$) and TLSs ($\Gamma_\text{t}$). As we show later, in our case the TLSs act as heat reservoir, because they provide the main relaxation channel for the qubit ($\Gamma_\text{q} \lesssim \sum_k \Gamma_\text{qt}^k$) while being approximately lossless ($\Gamma_\text{t} \approx 0$).
			\textbf{c} Schematic illustration of the qubit and TLS populations during the polarization sequence. Each cycle of the Szilard engine consists of a qubit preparation followed by the cross relaxation between the qubit and the TLSs. 
			After each cycle the polarization of the TLSs increases.
		} \label{fig:0} 
	\end{center}
\end{figure*}

Here, we implement a quantum Szilard engine~\cite{Szilard1929, Toyabe2010, Koski2014, Peterson2020} which manipulates the environment of a superconducting qubit. Our Szilard engine executes a hyperpolarization protocol similar to experiments using spin qubits~\cite{Bluhm2010} or defect centers~\cite{Broadway2018}, and is readily applicable in state of the art quantum processors~\cite{Corcoles2021Aug, Gold2021Sep, Satzinger2021Dec}.
The hyperpolarized environment reveals that the qubit is weakly coupled to a TLS environment of unknown origin, which relaxes over tens of milliseconds. Conversely, this heretofore hidden environment can now be identified to be the dominant loss mechanism of our qubit, and we dread that similarly acting environments are ubiquitous in superconducting hardware. The quantum Szilard engine consists of a granular aluminum fluxonium qubit~\cite{Gruenhaupt2019} that can be actively prepared in one of its eigenstates $|\text{g}\rangle$ or $|\text{e}\rangle$. The fluxonium and its complex environment is depicted in \frefadd{fig:0}{a}. The Szilard engine implements a dynamical polarization protocol on the TLSs as illustrated (\frefadd{fig:0}{b,c}). In contrast to nuclear hyperpolarization, here the qubit and the TLSs operate in the same frequency domain, requiring active or autonomous feedback schemes.

\begin{figure}[ht!]
	\begin{center}
		\includegraphics[scale=1.0]{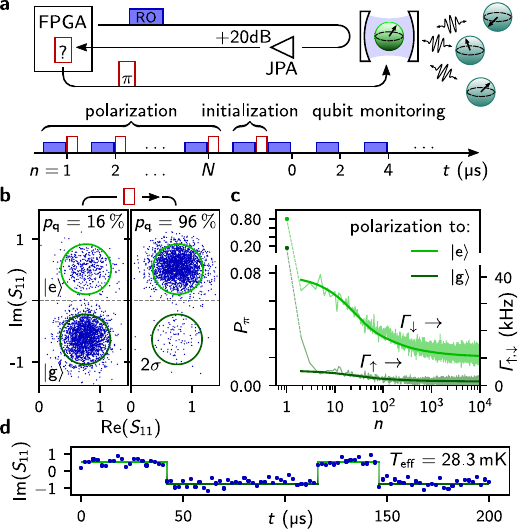}
		\caption{
			\textbf{Szilard engine in action.}
			\textbf{a} Schematic drawing of the experiment and the control sequence implementing a Szilard engine. The qubit consists of a fluxonium biased at half flux (Supp.~\ref{app:fluxonium}) operating at with its fundamental transition $f_{01} = \SI{1.2}{GHz}$ separated by \SI{6.6}{GHz} from the higher levels. The qubit is coupled to an unknown mesoscopic environment which, as we will show in \fref{fig:2}~and~\fref{fig:3} can be modeled as an ensemble of TLSs. We start the experiment with the TLS polarization sequence (\frefadd{fig:0}{c}) by stabilizing the qubit to either $|\text{g}\rangle$ or $|\text{e}\rangle$ using $N$ active feedback preparations.
			This is followed by a qubit initialization to $|\text{g}\rangle$ or $|\text{e}\rangle$, and immediately after we begin to monitor the qubit state stroboscopically. For the polarization and the qubit monitoring the repetition time is $t_\text{rep} = \SI{2}{\micro s}$, much shorter than the qubit's relaxation time $T_1 \approx \SI{20}{\micro s}$. Before each of the 2500 repetitions, we wait for \SI{50}{ms} to allow the environment to relax. The protocol is orchestrated by the field programmable gate array (FPGA) controller from Quantum Machines, with an internal real-time feedback latency of $\sim\!\SI{200}{ns}$ (Supp.~\ref{app:setup} provides a schematic of the detailed setup).
			\textbf{b} Scatter plot of the complex reflection coefficient $S_{11}$ of the readout signal for the qubit in equilibrium (left panel) and after $|\text{e}\rangle$-state preparation (right panel). The readout integration time is \SI{128}{ns} resulting in a separation of $5.6\,\sigma$ (green circles indicate $2\,\sigma$). \textbf{c} Probability $P_{\pi}$ to reset the qubit to its target state during the polarization ($P_\pi$ is corrected for state preparation and measurement errors, Supp.~\ref{app:pi_histogram}). Using $t_\text{rep}$, the values of $P_{\pi}$ can be mapped to the qubit transition rates $\Gamma_\uparrow$ and $\Gamma_\downarrow$ for polarization to $|\text{g}\rangle$ and $|\text{e}\rangle$, respectively (right hand axis). The evolution of the rates is captured by the theoretical model derived in the main text (solid lines).
			\textbf{d} Typical quantum jump trace during qubit monitoring (as shown in \textbf{a}). The solid line indicates the assigned qubit state.
		} \label{fig:1}
	\end{center}
\end{figure}

The experimental workflow (\frefadd{fig:1}{a}) starts with a polarization sequence where we stabilize the qubit in either $|\text{g}\rangle$ or $|\text{e}\rangle$, thereby cooling or heating the reservoir, respectively. After polarizing the reservoir, the qubit is initialized to $|\text{g}\rangle$ or $|\text{e}\rangle$ and the combined qubit and reservoir system relaxes to its steady state. As an example (\frefadd{fig:1}{b}), we show the qubit population before and after the first preparation in a sequence polarizing to $|\text{e}\rangle$.
The amount of heat in the reservoir, i.e. the degree of TLS polarization, varies with the operation time of the Szilard engine, given by the number of qubit preparations~$N$. Correspondingly, in \frefadd{fig:1}{c}, we show the measured decrease of qubit transition rates $\Gamma_{\uparrow,\downarrow}$ during stabilization in $|\text{g}\rangle$ or $|\text{e}\rangle$, respectively.
The different polarization and initialization scenarios are measured interleaved with $M=2500$ repetitions for each scenario, which sets the uncertainties visible as noise in the measured curves.

\begin{figure*}[ht!]
	\begin{center}
		\includegraphics[scale=1.0]{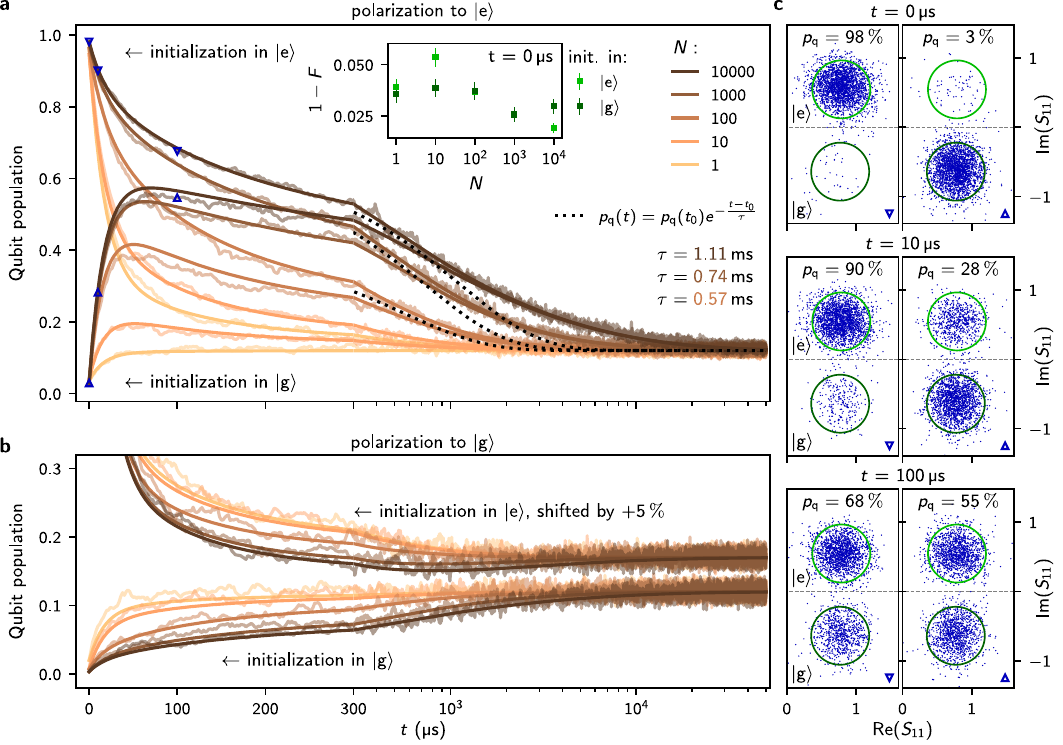}
		\caption{\textbf{Qubit evolution after running the Szilard engine.} \textbf{a} Measured relaxation of the qubit after polarization to $|\text{e}\rangle$ for various times $N \cdot t_\text{rep}$ followed by an initialization to either $|\text{g}\rangle$ or $|\text{e}\rangle$. Note the logarithmic $x$-axis from \SI{300}{\micro s} onwards, which is required to depict the slow relaxation dynamics. The exponential decay curves shown in dotted lines, with the decay times indicated by the corresponding labels, are guides to the eye to illustrate the non-exponential relaxation of the environment (Supp.~\ref{app:tails}). The inset shows the preparation infidelity of the initialization. We observe an increasing fidelity with $N$, in particular for the initialization in $|\text{e}\rangle$. The errorbars show the one sigma confidence intervals of the binomial distribution with 2500 repetitions.
		\textbf{b} Measured relaxation of the qubit after polarization to $|\text{g}\rangle$ followed by an initialization to either $|\text{g}\rangle$ or $|\text{e}\rangle$. Compared to panel a, the opposite effect is visible: the environment is cooled by the polarization sequence, demonstrating that the heat flow in the environment is not the trivial result of heating due to repeated microwave readout and control pulses. The upper curves are shifted upwards by \SI{5}{\percent} for better visibility. The continuous lines in panel a and b correspond to the theoretical model of \eqsref{eq:rate_eq_qubit}~and~\ref{eq:rate_eq_tls}, applied to all measured curves simultaneously. 
		\textbf{c} Scatter plots of the complex reflection coefficient $S_{11}$ for the relaxation curves shown in panel a for $N=10^4$. The left panels illustrate the reduced relaxation of the excited state population vs. time. The right panels demonstrate that the qubit undergoes a population inversion due to interactions with the environment. Notably, the $|\text{f}\rangle$-state is not populated, as illustrated by the absence of a third cloud in the $S_{11}$ distribution (Supp.~\ref{app:high_temp}).
	} \label{fig:2}
	\end{center}
\end{figure*}

\begin{figure*}[ht!]
	\begin{center}
		\includegraphics[scale=1.0]{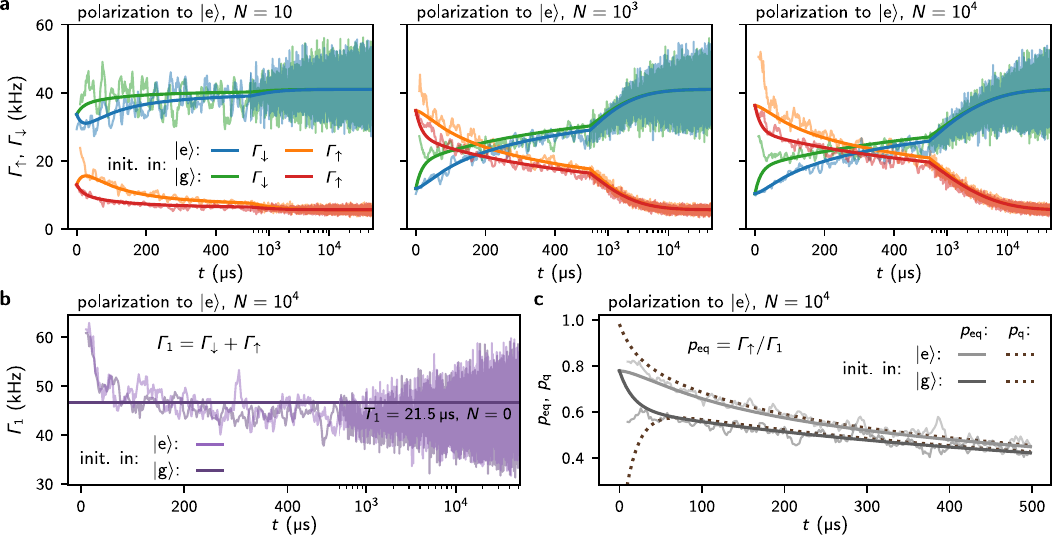}
		\caption{\textbf{Measured population inversion and constant relaxation rate, signatures of the TLS environment.} \textbf{a} 
			Measured (in light color) and calculated (in dark color) qubit transitions rates $\Gamma_{\uparrow,\downarrow}$ following initialization to $|\text{g}\rangle$ or $|\text{e}\rangle$ and for increasing polarization to $|\text{e}\rangle$ from $N=10,\,10^3\,\text{to}\,10^4$ shown in the left, center and right panel, respectively. The measured rates are extracted from the same quantum jump traces used to extract the qubit relaxation (\frefadd{fig:2}{a}), with the logarithmic time-axis starting at \SI{500}{\micro s}. For long polarization times the rates are reversed in the beginning, meaning that the  qubit sees a negative temperature environment. Note that in all cases the $|\text{g}\rangle$ state initialization visibly cools the environment, suggesting a  heat capacity of only a few energy quanta.
			In order to reduce the statistical noise, a five-point  moving average filter was applied corresponding to a \SI{10}{\micro s} window. Furthermore, the first \SI{10}{\micro s} of the orange and green curve are omitted due to the low statistics; it is unlikely to have two consecutive $\lesssim\SI{5}{\micro s}$ intervals between jumps. Similarly, these rates are overestimated in the beginning by preferably detecting short $T_1$ fluctuations of the qubit.
			\textbf{b} Relaxation time $\Gamma_1$ obtained from the $\Gamma_{\uparrow, \downarrow}$ rates in panel a for $N=10^4$. The $\Gamma_1$ rate is constant compared to the changes in $\Gamma_\uparrow$ and $\Gamma_\downarrow$ shown in panel a, i.e. $\Gamma_1$ is independent of the environmental populations, which indicates a TLS environment. \textbf{c} Equilibrium population of the qubit $p_\text{eq}$ extracted from the $\Gamma_{\uparrow, \downarrow}$ rates. The dashed lines show the corresponding qubit population $p_\text{q}$ relaxation taken from \frefadd{fig:2}{a}. We extrapolate an effective population of the environment $p_\text{eq} = \SI{78}{\percent}$ at $t = 0$. In all panels, the theoretical curves use the same parameters as that in \fref{fig:2}.
		} \label{fig:3}
		\vspace*{-0.25cm}
	\end{center}
\end{figure*}

The relaxation of the reservoir can not be directly observed and has to be inferred from the qubit dynamics. While the common approach is to measure the free decay of the qubit (Supp.~\ref{app:free_decay}), here we exploit the fact that the qubit readout is more than \SI{96}{\percent} quantum non-demolishing (Supp.~\ref{app:qnd} and Ref.~\cite{Gusenkova2021}) and we perform repeated single shot readouts, resulting in stroboscopic quantum jump traces~(\frefadd{fig:1}{d}). The main benefit of this method is the direct determination of the transition rates $\Gamma_{\uparrow,\downarrow}$ between the ground and excited state, which allows us to discriminate between changes in the energy relaxation rate and changes in the equilibrium population of the qubit. In \fref{fig:2} we show measured qubit relaxation curves for several polarization and initialization scenarios. 
Note that for long enough polarization times to the excited state ($N \geq 10^3$) the qubit reaches population inversion (\frefadd{fig:2}{c}, bottom), which hints at a population inversion of the reservoir. This effect is also confirmed by the inversion of the transition rates $\Gamma_\uparrow > \Gamma_\downarrow$ (\frefadd{fig:3}{a}). A notable consequence is that for $N = 10^4$ the preparation fidelity for the excited state is higher than for the ground state (\frefadd{fig:2}{a}, inset).

The time-evolving transition rates (\frefadd{fig:3}{a}) are obtained from the stroboscopic quantum jump traces (\frefadd{fig:1}{d}) by using $\Gamma_\uparrow = - \ln (P_{|\text{g}\rangle, |\text{g}\rangle}) / t_\text{rep}$ and $\Gamma_\downarrow = - \ln(P_{|\text{e}\rangle, |\text{e}\rangle}) / t_\text{rep}$, where $P$ is the probability to measure the same qubit state in successive measurements, and $t_\text{rep}$ is the repetition time (Supp.~\ref{app:rates}). These rates define the relaxation rate  $\Gamma_1 = \Gamma_\uparrow + \Gamma_\downarrow$ and the equilibrium population of the qubit $p_\text{eq} = \Gamma_\uparrow / \Gamma_1$. 
Note that the noise magnitude varies with the qubit population (\frefadd{fig:3}{a}), because the rates $\Gamma_{\uparrow,\downarrow}$ are based on conditional probabilities.
Remarkably, after a heating sequence with $N=10^4$, $\Gamma_1$ of the qubit is comparably constant (\frefadd{fig:3}{b}); in contrast $p_\text{eq}$ follows a non-exponential relaxation for time scales up to \SI{50}{ms}. At the end of the polarization sequence we can ascribe for the TLSs a hyperpolarization $p_\text{eq}^\text{TLSs} = \SI{97}{\percent}$, which, when taking into account the intrinsic loss of the qubit, gives the measured $p_\text{eq} = \SI{78}{\percent}$ (\frefadd{fig:3}{c}).
Conversely, after a cooling sequence with $N=10^4$, we extract $p_\text{eq} = \SI{3.0}{\percent}$, as can be ascertained in \frefadd{fig:2}{b} using that the qubit population $p_\text{q} = \SI{2.0}{\percent} \approx p_\text{eq}$ after $1/\Gamma_1$. 
Hence, the Szilard engine cooled the environment to an effective temperature of \SI{16}{mK}, which is well below the temperature of the dilution refrigerator $\sim \SI{25}{mK}$ and the effective temperature $T_{\text{eff}}=\SI{28.3}{mK}$ corresponding to the idle qubit population $p_\text{th} = \SI{12,0}{\percent}$ (\frefadd{fig:1}{d}). The TLS hyperpolarization is even lower, $p_\text{eq}^\text{TLSs} = 3.4\,\text{\textperthousand}\,\hat{=}\,\SI{9.9}{mK}$, limited by the qubit preparation infidelity. The values are extrapolated from the theoretical model, which will be explained in the next paragraph. For both, heating and cooling, the hyperpolarization values are among the highest reported in literature \cite{Yang2009Jun, Wang2013}.

The constant relaxation rate $\Gamma_1$ as well as the observed population inversion indicate an environment consisting of TLSs. We therefore model the system assuming the qubit to be coupled to a countable number of TLSs with populations ${p}_\text{t}^k$.
The cross relaxation rates $\Gamma_\text{qt}^k$ between the qubit and the TLSs are given by~\cite{Solomon1955, Barends2013}
\begin{align}
	\Gamma_\text{qt}^k = \frac{2g^2\Gamma_2}{\Gamma_2^2 + \delta_k^2}, \label{eq:gamma_qts}
\end{align}
where $\delta_k$ is the detuning between the qubit and the $k^\text{th}$~TLS, $g$ their transverse coupling strength, and $\Gamma_2$ the sum of their decoherence rates. Since the TLSs can in turn excite the qubit, we conclude that the qubit and the TLSs are close in frequency so that they relax approximately to the same thermal population $p_\text{th}$ (note that the qubit is well thermalized, as discussed earlier).
Finally, we introduce intrinsic relaxation rates for the qubit and the TLSs, $\Gamma_\text{q}$ and $\Gamma_\text{t}^k$, respectively, capturing the remaining environment (\frefadd{fig:0}{b}).
The dynamics is governed by the so-called Solomon equations~\cite{Solomon1955}, extensively used in the field of nuclear hyperpolarization \cite{Voegli2014}. The rate equations read:
\begin{align}
	\dot{p}_\text{q} &=  -\Gamma_\text{q} (p_\text{q} - p_\text{th}) - \sum_k\Gamma_\text{qt}^k (p_\text{q} - p_\text{t}^k)  \label{eq:rate_eq_qubit}\\
	\dot{p}_\text{t}^k &= - \Gamma_\text{t}^k (p_\text{t}^k - p_\text{th}) - \Gamma_\text{qt}^k (p_\text{t}^k - p_\text{q}), \label{eq:rate_eq_tls}
\end{align}
where we identify the constant qubit relaxation rate $\Gamma_1 = \Gamma_\text{q} + \sum_k \Gamma_\text{qt}^k$ and the time-dependent \linebreak $p_\text{eq} = \left(\Gamma_\text{q}p_\text{th} + \sum_k \Gamma_\text{qt}^k p_\text{t}^k \right)/\, \Gamma_1$. As a consequence of \eref{eq:rate_eq_tls}, during the polarization time $N \cdot t_\text{rep}$, when we enforce $p_\text{q} = 0$~or~$1$, there is an exponential population transfer between the qubit and each TLS, and at the end of the sequence we expect to find the TLSs polarized (\frefadd{fig:1}{c}).

So far, the model in \eqsref{eq:rate_eq_qubit}~and~\ref{eq:rate_eq_tls} requires two rates for each TLS. In order to extract meaningful information from the measurements by virtue of \eref{eq:gamma_qts} we need to make simplifying assumptions and reduce the number of fitting parameters. Since we observe the TLS polarisation in different qubits and at different qubit frequencies (Supp.~\ref{app:further_TLS_observations}) we expect the TLSs to be randomly distributed in frequency. We simplify this distribution by modeling them to be equally spaced in frequency with $\delta_k = k\Delta + \Delta_0$, where $\Delta_0 \in [0, \Delta /2]$ defines a shift of the TLS ladder with respect to the qubit frequency. This is justified by the fact that we are mainly interested in capturing the slow, non-exponential relaxation at millisecond timescales. With the same argument, we assume for all TLSs the same $g$ and $\Gamma_2$. The price we pay using these simplification is that the model captures less accurately the initial features of the decay curves, at $t < \SI{300}{\micro s}$. Indeed, these features are a fingerprint of the exact configuration of the TLSs, and, as expected, they fluctuate in time \cite{Klimov2018Aug,Thorbeck2022Oct} (Supp.~\ref{app:free_decay}).

The simplified model allows to rewrite \eref{eq:gamma_qts} in the compact form $\Gamma_\text{qt}^k = a b^2 / [b^2 + (k + b c)^2]$, showing that $g$, $\Delta$, $\Delta_0$ and $\Gamma_2$ do not appear independently in the model. Instead, $g = \sqrt{a \Gamma_2 / 2}$, $\Delta = \Gamma_2 / b$ and $\Delta_0 = c \Gamma_2$ can be determined for a given decoherence rate from a successful fit of the model.
The fit procedure is further restricted by inserting the measured qubit relaxation rate $\Gamma_1 = 1 / \SI{21.5}{\micro s}$ (\frefadd{fig:3}{b}), leaving us with only two essential fit parameters $\Gamma_\text{q}$ and $b$ (Supp.~\ref{app:fitting}). 
The robustness of the model is illustrated by the fact that a fit of only the first millisecond to one of the stronger polarized relaxation curves (e.g. polarization to $|\text{e}\rangle$ for $N=10^3$ with initialization in $|\text{g}\rangle$ or $|\text{e}\rangle$) is sufficient to describe the highly non-exponential relaxation of all measurements on the entire relaxation range up to \SI{50}{ms} (\fref{fig:2} and \fref{fig:3}, continuous lines). Details of the fitting procedure are presented in Supp.~\ref{app:fitting}.

Using the lower bound $\Gamma_2 \geq \Gamma_2^\text{q} \approx \SI{0.5}{MHz}$, where $\Gamma_2^\text{q}$ is the decoherence rate of the qubit (Supp.~\ref{app:coherence}), we extract $g \geq 2 \pi \cdot \SI{12}{kHz}$ and $\Delta \geq 2\pi \cdot \SI{167}{kHz}$. The comparably small coupling strength $g \ll \Gamma_2^\text{q}$ is consistent with the fact that we do not observe avoided level crossings in the qubit spectrum. Notably, this argument remains valid even for higher decoherence because $g$ and $\Delta$ scale with $\sqrt{\Gamma_2}$ and $\Gamma_2$, respectively. 
Using an upper bound for the decoherence $\Gamma_2 \sim 1 / \SI{10}{ns} \ll f_\text{q}$, comparable to values reported in Ref.~\cite{Lisenfeld2019}, gives $g < 2\pi \cdot \SI{170}{kHz} $ and ${\Delta < 2\pi \cdot \SI{35}{MH}}$.

Furthermore, we can calculate the two contributions of the qubit relaxation: one rate is due to interactions with the TLSs, $\Gamma_\text{qt}^\text{TLSs} = \sum_k \Gamma_\text{qt}^k= \SI{35.9}{kHz}$, and the other is the remaining intrinsic relaxation $\Gamma_\text{q} = \SI{10.7}{kHz}$. We therefore identify the TLS bath as the dominant loss mechanism. Remarkably, the fit also indicates that the intrinsic relaxation time exceeds $1/\Gamma_\text{t} \geq \SI{50}{ms}$, which is orders of magnitude longer than previously measured relaxation rates of dielectric TLSs~\cite{Neeley2008, Lisenfeld2010, Lisenfeld2016}. This fact leads us to believe that we are reporting a new type of TLS environment, possibly related to spins \cite{Lee1977Feb, Yang2020Sep} or trapped quasiparticle TLSs~\cite{Graaf2020}. Finally, we would like to mention that $\Gamma_\text{qt}^k \geq \Gamma_\text{t}$ for $|k| \leq 15$, which means that the qubit is the main decay channel for at least the first few tens most resonant TLSs.

Following Szilard's seminal paper~\cite{Szilard1929}, the homonymous engine uses measured information as fuel (Supp.~\ref{app:szilard}). In the first iteration of a cooling sequence starting from thermal equilibrium $T= \SI{28.3}{mK}$, the engine extracts on average the internal energy $\Delta U = 0.24\,k_\text{B} T$ from the qubit,  corresponding to an entropy reduction of $0.37\,k_\text{B}$, which should be compared with the entropy produced by the measurement apparatus $k_\text{B}\ln 2 \approx 0.69\,k_\text{B}$. From the rate equation we can calculate the optimal working regime for our Szilard engine. Using the fitted parameters we infer that the maximum heat reduction $\Delta Q = 0.11\,k_\text{B} T$ in the reservoir occurs $\SI{68}{\micro s}$ after the qubit initialization. Thus, at most half of the extracted heat from the qubit can be used to cool the reservoir. With a similar timescale of $t_\text{rep} = \SI{100}{\micro s}$ we show in Supp.~\ref{app:pi_pulse_heating} that the reservoir can also be heated by a sequence of $\pi$-pulses. However, this procedure, introduced in Ref.~\cite{Gustavsson2016Dec}, can not result in a population inversion in the reservoir.

In summary, using a superconducting qubit and active feedback we demonstrated a quantum Szilard engine, which can polarize a TLS environment of unknown origin. As a result, the qubit's population exhibits remarkably long and non-exponential dynamics due to the intrinsically long decay time of the TLSs, exceeding \SI{50}{ms}. This showcases the challenges and pitfalls of extracting $T_1$ from qubit population relaxation data. In our device, we extract $T_1$ from quantum jumps and show that it is unaffected by continuous operation of the qubit, ruling out enhanced quasiparticle diffusion \cite{Gustavsson2016Dec}. While $T_1$ is independent of the environment population, the transition rates $\Gamma_{\uparrow,\downarrow}$ are not. Our results are particularly relevant in the context of quantum processors, where the heating and cooling of the environment is a byproduct of continuous operation. The Szilard engine could be used to study out-of-equilibrium processes or to preferentially reduce one of the qubit transition rates. For example, reducing $\Gamma_{\uparrow}$ would be beneficial for bosonic codes~\cite{Reinhold2020Aug, Grimm2020Aug}. 

In our system quantum coherence between the qubit and the TLSs can be neglected, allowing a simple description using the Solomon equations. As quantum hardware continues to improve, coherent interactions and non-Markovian qubit dynamics will start to play a role, raising the bar for quantum error correction strategies. The quantum Szilard engine presented here offers a first glimpse of the challenges facing future hardware, in which coherence improvements also translate into increasingly complex interactions with the environment.
\vspace*{0.4cm} 
\begin{center}
	\bf Acknowledgements
\end{center}
We are grateful to J. Lisenfeld and W. Wulfhekel for insightful discussions and to A. Lukashenko and L. Radtke for technical assistance. Funding was provided by the Alexander von Humboldt Foundation in the framework of a Sofja Kovalevskaja award endowed by the German Federal Ministry of Education and Research, and by the European Union's Horizon 2020 programme under No. 899561 (AVaQus). M.S. and G.C. acknowledge support from the German Ministry of Education and Research (BMBF) within the project GEQCOS (FKZ: 13N15683 and 13N15685). P.P. acknowledges support from the German Ministry of Education and Research (BMBF) within the QUANTERA project SiUCs (FKZ: 13N15209). D.R., S.G. and W.W. acknowledge support by the European Research Council advanced grant MoQuOS (no. 741276). Facilities use was supported by the KIT Nanostructure Service Laboratory. We acknowledge qKit for providing a convenient measurement software framework.
\vspace{0.1cm}
\begin{center}
	\bf Methods \nopagebreak
\end{center}
\nopagebreak
\noindent Supp.~\ref{app:fluxonium} provides details on the fluxonium qubit and Supp.~\ref{app:setup} provides details on the microwave setup.
\begin{center}
	\bf Data availability
\end{center}
\noindent Raw and processed data are publicly available via Zenodo at \url{https://doi.org/10.5281/zenodo.7817552}. Additional data are available from the corresponding authors upon reasonable request.
\vspace*{0.1cm}
\begin{center}
	\bf Code availability
\end{center}
\noindent The analysis script used to generate the data in the figures is publicly available via Zenodo at \url{https://doi.org/10.5281/zenodo.7817552}.
\vspace*{0.1cm}
\begin{center}
	\bf Author contributions
\end{center}
\noindent M.S. and I.M.P. conceived the experiment. N.D. and Y.C. installed and supervised the real-time microwave electronics. M.S. performed the experiment, analysed the data and developed the theoretical model. I.T. and P.W. designed and fabricated the parametric amplifier. R.G. and O.S. provided the real-time microwave electronics for preliminary experiments. P.P. and N.G. carried out additional measurements for the revised manuscript. M.S. and I.M.P. wrote the manuscript. A.S. and I.M.P. supervised the project. All authors discussed the results and contributed to the final manuscript.
\vspace*{0.1cm}
\begin{center}
	\bf Additional information
\end{center}
\noindent \textbf{Correspondence and requests} for materials should be addressed to Martin Spiecker or Ioan M. Pop.\\

\noindent \textbf{Peer review information} Nature Physics thanks Tobias Lindström, Nathan Earnest-Noble and the other, anonymous, reviewer(s) for their contribution to the peer review of this work.\\

\noindent \textbf{Reprints and permissions information} is available at \url{www.nature.com/reprints}.

\addtocontents{toc}{\SkipTocEntry}
\bibliography{Bibliography}


\clearpage
\onecolumngrid

\begin{center}
	\large \bf Supplementary information for\\Two-level system hyperpolarization using a quantum Szilard engine
\end{center}

\begin{center}
	Martin Spiecker,$^{1,2}$ Patrick Paluch,$^{1,2}$ Nicolas Gosling,$^2$ Niv Drucker,$^3$
	Shlomi Matityahu,$^4$\\
	Daria Gusenkova,$^{1,2}$ Simon Günzler,$^{1,2}$ Dennis Rieger,$^{1,2}$
	Ivan Takmakov,$^{1,2}$ Francesco Valenti,$^2$\\
	Patrick Winkel,$^{1,2}$ Richard Gebauer,$^5$ Oliver Sander,$^5$ Gianluigi Catelani,$^6$ Alexander Shnirman,$^{2,4}$\\
	Alexey V. Ustinov,$^{1,2}$ Wolfgang Wernsdorfer,$^{1,2}$ Yonatan Cohen,$^3$ and Ioan M. Pop$^{1,2}$\\[0.1cm]
	\small
	\textit{$^1$PHI,~Karlsruhe~Institute~of~Technology,~76131~Karlsruhe,~Germany\\
	$^2$IQMT,~Karlsruhe~Institute~of~Technology,~76344~Eggenstein-Leopoldshafen,~Germany\\
	$^3$Quantum Machines, 67443 Tel Aviv-Yafo, Israel\\
	$^4$TKM,~Karlsruhe~Institute~of~Technology,~76131~Karlsruhe,~Germany\\
	$^5$IPE,~Karlsruhe~Institute~of~Technology,~76344~Eggenstein-Leopoldshafen,~Germany\\
	$^6$JARA,~Forschungszentrum Jülich, 52425 Jülich, Germany}\\
	(Dated: \today)
\end{center}

\setcounter{section}{0}
\setcounter{figure}{0}
\setcounter{equation}{0}
\renewcommand{\thesection}{\Alph{section}}
\renewcommand{\thefigure}{S\arabic{figure}}
\renewcommand{\theequation}{S\arabic{equation}}

\tableofcontents

\newpage
\section{Free decay of the qubit after TLS polarization.} \label{app:free_decay}
\begin{figure}[H]
\begin{center}
\includegraphics[scale=1.0]{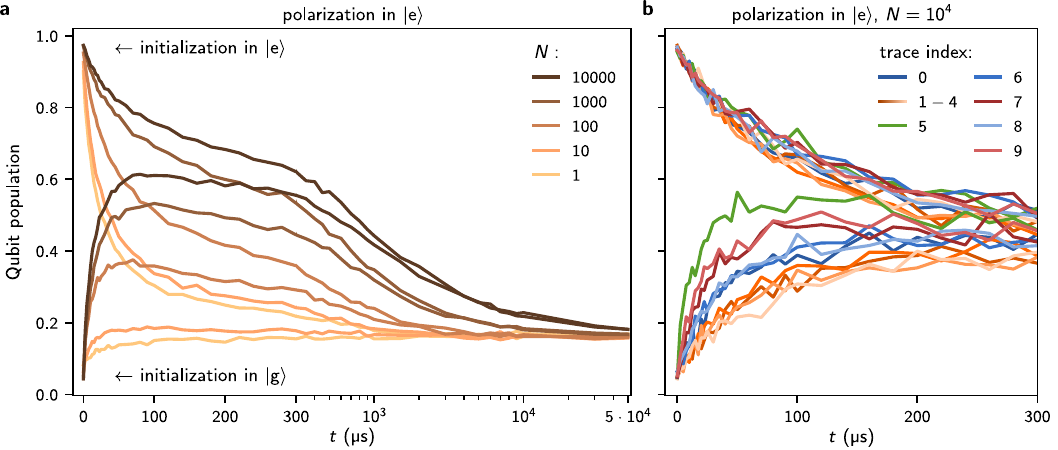}
\caption{\textbf{Measured free decay after polarization to $|\text{e}\rangle$.} \textbf{a} Free decay of the qubit  after polarization for various times $N \cdot t_\text{rep}$ followed by an initialization to either $|\text{g} \rangle$ or $|\text{e}\rangle$. Similarly to \frefadd{fig:2}{a} in the main text, the qubit reaches a population inversion of approximately $\SI{60}{\percent}$ after being initialized in $|\text{g}\rangle$.
\textbf{b} We observe fluctuations at the beginning of the relaxation curves after polarization to $|\text{e}\rangle$, especially for initialization in $|\text{g}\rangle$. The color coding is used to indicate quantitatively different time evolutions, which can be grouped in at least four families (blue, orange, green and red), suggesting at least four different configurations of the TLS environment. Following \eref{eq:gamma_qts}, the main contributions can be expected from the most resonant TLSs that affect in particular the beginning of the relaxation curves.
A corresponding behaviour for each trace index is also observed for the curves after polarization to $|\text{g}\rangle$, which are measured interleaved (not shown), in which case the fluctuations are mainly seen for the initialization in $|\text{e}\rangle$. The data set shown in the left panel was selected to be one of the few without fluctuations. The time interval between measurements with successive indices is $\sim\SI{4}{h}$. These random fluctuations are also present in data obtained from quantum jumps. Nevertheless, since the data acquisition time is more than ten times faster for quantum jumps experiments, it is less likely to observe such fluctuations. This allows us to fit simultaneously the data sets for polarization to $|\text{g} \rangle$ and $|\text{e}\rangle$ for several values of $N$, as shown by the continuous lines in main text \fref{fig:2} and \fref{fig:3}.} \label{fig:free_decay} 
\end{center}
\end{figure}

In \frefadd{fig:free_decay}{a} we show the measured qubit free decay after polarization to $|\text{e}\rangle$. We observe qualitatively the same behavior as in \frefadd{fig:2}{a} in the main text, which was measured with quantum jumps. The whole set of experiments including the four polarization and initialization scenarios each with various $N$ values lasts for approximately $\SI{40}{h}$ and can be affected by drifts in the environment, as illustrated in \frefadd{fig:free_decay}{b}.

\newpage
\section{Qubit relaxation as a function of the readout repetition time} \label{app:qnd}

\begin{figure}[H]
\begin{center}
\includegraphics[scale=1.0]{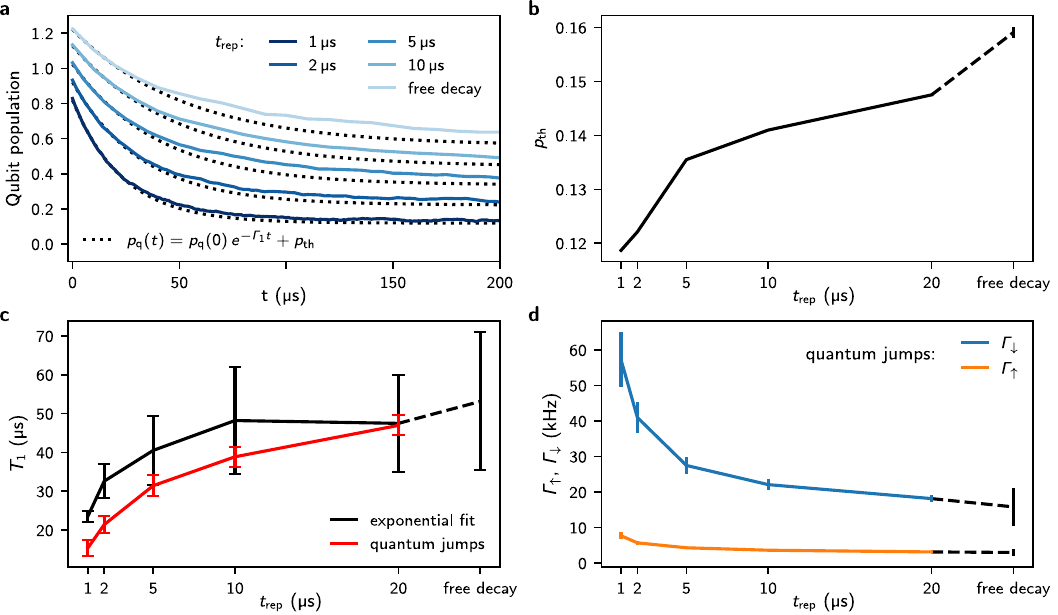}
\caption{\textbf{Quantum demolition effects induced by the readout.} \textbf{a} Qubit relaxation obtained from quantum jump traces for various repetition times $t_\text{rep}$ and from free decay. The curves are successively shifted vertically by $\SI{10}{\percent}$ for visibility. The deviation from the exponential function (dotted line), especially close to the thermal equilibrium, is an indication that the TLS reservoir was heated by the energy deposited in the qubit during the initial $\pi$-pulse (\fref{fig:pi_pulse_heating}). \textbf{b} The shift of the thermal qubit population $p_\text{th}$ to lower values with decreasing repetition time indicates a cooling effect of the measurement on the qubit. When not visible the error bars are in the range of the line width. \textbf{c} Relaxation times $T_1$ extracted either from exponential fits to the data in panel a or from quantum jumps. 
For the exponential fits we use the first \SI{20}{\micro s} of the decay together with the baseline $p_\text{th}$ (panel b) to extract the $T_1$ values (for the deviation see discussion in the text).
In both cases we observe shorter $T_1$ values at faster repetition times, indicating the increasing quantum demolition contribution to the energy relaxation budget. The results reported in the main text were measured with $t_\text{rep} = \SI{2}{\micro s}$, for which approximately half of the relaxation can be attributed to the quantum demolishing of the readout.  
\textbf{d} Measured qubit transition rates extracted from quantum jump traces for various $t_\text{rep}$. The quantum demolishing of the readout mainly consists in an increased $\Gamma_\downarrow$-rate. In all panels the lines connecting the points are guides to the eye. In all panels the error bars correspond to the standard deviation.
} \label{fig:qnd_effects}
\end{center}
\end{figure}

First, we discuss the influence of the measurement on the qubit relaxation for which we contrast different $T_1$-experiments in \frefadd{fig:qnd_effects}{a}. The measurement increases both transition rates $\Gamma_{\uparrow,\downarrow}$ compared to the free decay rates (\frefadd{fig:qnd_effects}{d}). However, for the relaxation rate $\Gamma_1 = \Gamma_\uparrow + \Gamma_\downarrow$ the $\Gamma_\downarrow$-rate contributes dominantly. 
The probability per measurement to decay from the excited state is approximately \SI{4}{\percent}, which, for a repetition time $t_\text{rep}=\SI{2}{\micro s}$, corresponds to an additional rate of $\Gamma_\downarrow^\text{M}\approx\SI{20}{kHz}$ induced by the measurement. The relative increase of $\Gamma_\downarrow$ exceeds the one of $\Gamma_\uparrow$ and therefore lowers the qubit's effective temperature compared to free decay (\frefadd{fig:qnd_effects}{b}).

Second, we illustrate the challenges in measuring the relaxation rate of a qubit coupled to a finite TLS environment. From the quantum jumps analysis, we obtain $T_1= \SI[separate-uncertainty]{21.4+-2.2}{\micro s}$
for $t_\text{rep} = \SI{2}{\micro s}$ (as reported in the main text). In contrast, an exponential fit to the data shown in \frefadd{fig:qnd_effects}{a} results in higher $T_1$ values even though we conservatively use only the first \SI{20}{\micro s} for the fit (\frefadd{fig:qnd_effects}{c}). This discrepancy is a consequence of the finite size of the TLS environment. The energy transferred into the environment by the initial qubit $\pi$-pulse is sufficient to create the illusion of an increased relaxation time (\frefadd{fig:pi_pulse_heating}{a}, right panel).
The difference illustrates the importance of the quantum jump method for measuring the energy relaxation. 
However, when $t_\text{rep}$ approaches $T_1$, the quantum jumps method also overestimates the relaxation time, as visible in \frefadd{fig:qnd_effects}{c} where the quantum jumps method approaches the free decay $T_1$-time extracted from the exponential fit for $t_\text{rep} = \SI{20}{\micro s}$.

\newpage
\section{The fluxonium artificial atom} \label{app:fluxonium}
The device under study is a fluxonium artificial atom that can be measured by the dispersive frequency shift of its inductively coupled readout resonator. Both the fluxonium circuit and the resonator exploit the high kinetic inductance of granular aluminum. 
The device is identical to the one investigated in Ref.~\cite{Gusenkova2021} and presented in the supplementary material of Ref.~\cite{Gruenhaupt2019}.
The Josephson junction of the fluxonium is realized with a superconducting quantum interference device (SQUID) to have a flux tunable Josephson energy. In the following, we discuss the fluxonium Hamiltonian and point out two consequences that emerge from the SQUID implementation. For a time-independent external flux bias the Hamiltonian can be transformed to read
\begin{align}
H &=  \frac{1}{2C} \, q^2 + \frac{1}{2L} \, \phi^2  - E_{\text{J}_1} \cos \left(\frac{2 \pi}{\Phi_0} \left(\phi - \Phi_\text{ext}^\text{l}\right)\right) \,-\, E_{\text{J}_2} \cos \left(\frac{2 \pi}{\Phi_0} \left(\phi - \Phi_\text{ext}^\text{l} - \Phi_\text{ext}^\text{s}\right)\right), \label{eq:fluxonium_hamiltonian}
\end{align}
where $q$ and $\phi$ are the charge and flux operators obeying the commutation relation $[q, \phi] = i \hbar$. The capacitance $C = \SI{6.9}{\femto F}$ is mainly formed by the capacitances of the Josephson junctions of the SQUID. The kinetic superinductance $L = \SI{231}{\nano H}$ and the inner Josephson junction with the Josephson energy $E_{\text{J}_1}$ enclose the external flux $\Phi_\text{ext}^\text{l}$. The second Josephson junction with the Josephson energy $E_{\text{J}_2}$ encloses an additional flux $\Phi_\text{ext}^\text{s}$ with the first junction, the external flux of the SQUID.
Introducing the dimensionless flux variable $\varphi$, the flux-dependent Josephson energies in \eref{eq:fluxonium_hamiltonian} can be rewritten as
\begin{align*}
E_{\text{J}_1} \cos &(\varphi - \varphi_\text{l}) \,+\, E_{\text{J}_2} \cos (\varphi - (\varphi_\text{l} + \varphi_\text{s})) \\
& = \text{sign} \left(E_\Sigma\left(\varphi_\text{s} \right) \right) \cdot \sqrt{E_\Sigma\left(\varphi_\text{s} \right)^2 + E_\Delta\left(\varphi_\text{s} \right)^2}\,\cdot\, \cos\left(\varphi - \varphi_\text{ext} - \arctan \left( \frac{E_\Delta\left(\varphi_\text{s} \right)}{E_\Sigma\left(\varphi_\text{s} \right)}\right)\right)\\
&= E_\text{J}^\text{eff}(\varphi_\text{s}) \cdot \cos\left(\varphi - \varphi_\text{ext}^\text{eff}(\varphi_\text{l}, \varphi_\text{s})\right),
\end{align*}
where the flux dependent energies are $E_\Sigma\left(\varphi_\text{s} \right) =  (E_{\text{J}_1} + E_{\text{J}_2}) \cos \left(\varphi_\text{s} / 2\right)$ and $E_\Delta \left(\varphi_\text{s}\right) = (E_{\text{J}_2} - E_{\text{J}_1}) \sin\left(\varphi_\text{s}/2\right)$ and the external flux is defined by $\varphi_\text{ext} = \varphi_\text{l} + \varphi_\text{s} / 2$, showing that the SQUID flux contributes half to the external flux bias of the fluxonium.
The resulting fluxonium Hamiltonian has an effective Josephson energy $E_\text{J}^\text{eff}$ that only depends on the external flux in the SQUID loop and is flux biased by $\varphi_\text{ext}^\text{eff}$, which includes a nonlinear phase shift term that can directly be seen in the spectrum (Fig.~1b in Ref.~\cite{Gusenkova2021} or Fig.~S2 in Ref.~\cite{Gruenhaupt2019}). The device was operated at $\Phi_\text{ext} = 21.48 \, \Phi_0$ giving $E_\text{J}^\text{eff} = \SI{5.6}{GHz}$ and $\varphi_\text{ext}^\text{eff} / 2\pi = 0.5$.\\ 

The SQUID junction design comes with two implications. Firstly, the tunable Josephson energy is susceptible to local flux noise, which in the case of our device constitutes the main decoherence mechanism.
Secondly, the condition for destructive quasiparticle interference at the Josephson junction, which decouples the qubit from quasiparticle interactions~\cite{Pop2014Apr, Glazman2021}, is not met in our device. This was one of the reasons why we wanted to investigate the hypothesized quasiparticle activation in Ref.~\cite{Gustavsson2016Dec}. For our qubit we disprove this effect as detailed in Supp.~\ref{app:pi_pulse_heating}.

\newpage
\section{The time-domain setup} \label{app:setup}
\begin{figure}[H]
\begin{center}
\includegraphics[scale=1]{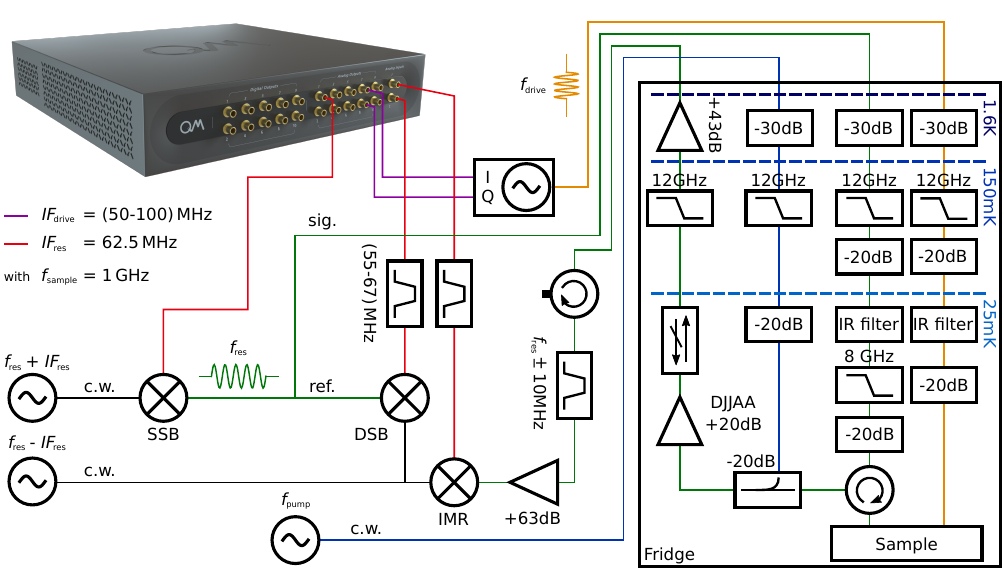}
\caption{\textbf{Experimental setup for qubit state measurements and active feedback.} 
The experimental workflow is orchestrated by the OPX instrument from Quantum Machines (visible in the top-left photograph). This FPGA-based instrument can be programmed to measure and estimate in real-time the qubit state using the $IQ$-demodulated readout signal, and it can trigger a $\pi$-pulse in order to prepare the qubit in its target state.
For the readout of the qubit, we use a \SI{128}{ns} long square pulse and an interferometric setup to purify the two-channel microwave generators that operate in continuous wave (c.w.) mode. The signal and the reference are interfered computationally by the OPX to extract the $I$ and $Q$ quadratures. We keep the intermediate frequency (IF) of the readout fixed at \SI{62.5}{MHz} yielding 16 samples per period for the integration.
The channels use different frequencies to account for the mixers operating on opposite sidebands.
For the qubit manipulation, we use a Gaussian envelope pulse with $\sigma = \SI{10}{ns}$ in a \SI{48}{ns} window.
\textbf{Fridge setup:} All microwave lines going into the cryostat are attenuated and filtered. A home made infrared (IR) filter employing Stycast\textsuperscript{\textregistered} ensures an attenuation of more than $\SI{-10}{\decibel}$ for frequencies larger than \SI{60}{\giga\hertz}.
On the way back the readout signal is first amplified with a home-made Dimer Josephson Junction Array Amplifier (DJJAA~\cite{Winkel2020Feb} providing $+\SI{20}{dB}$ of gain). We feed the pump tone for the DJJAA to the readout line with a directional coupler providing minimal loss for the readout signal. After the DJJAA the readout signal passes a \SI{40}{dB} isolation before it is further amplified by a high-electron-mobility transistor (HEMT). At room temperature the readout signal is routed through a home-made tunable filter in order to suppress the DJJAA pump tone, after which it is further amplified, down-converted to the intermediate frequency and finally recorded.
} \label{fig:setup}
\end{center}
\end{figure}

In \fref{fig:setup} we show a schematic of the microwave electronics setup for the fluxonium measurement and manipulation. 

\newpage
\section{Polarization of the environment while running the Szilard engine and SPAM errors} \label{app:pi_histogram}

During the polarization sequence one can observe a decrease in active feedback preparations of the qubit (\frefadd{fig:1}{c}). We quantify the probability to play a $\pi$-pulse on the qubit by $P'_\pi$, with prime denoting the measured probability containing state preparation and measurement (SPAM) errors. In the following we show how to extract the portion $P_\pi$ (without prime) which originates solely from qubit relaxation. 

The evolution of $P_\pi$ during the polarization sequence is explained by the TLS environment that becomes increasingly more polarized. The probability to measure the qubit relaxed from its target state during the repetition time $t_\text{rep}$ is
\begin{align}
   P_\pi(t) = 
    \left\{\begin{alignedat}{3}
      [1 - p_\text{eq}(t)] - [1 - p_\text{eq}(t)] e^{-\Gamma_1 t_\text{rep}},&\quad &\text{for polarization to $|\text{e}\rangle$},\\
     [p_\text{eq}(t) - 0] - [p_\text{eq}(t) - 0] e^{-\Gamma_1 t_\text{rep}}, &&\text{for polarization to $|\text{g}\rangle$}.
    \end{alignedat}\right.  \label{eq:Ppi_exact}
\end{align}
Since $\Gamma_1$ is constant (s. main text) the polarization of the TLSs is entirely encoded in $p_\text{eq}(t)$.  Note that the first term in each equation gives the probability to require a $\pi$-pulse starting from equilibrium. Using $\Gamma_1 = \Gamma_\uparrow + \Gamma_\downarrow$ and $p_\text{eq} = \Gamma_\uparrow / \Gamma_1$, \eref{eq:Ppi_exact} can be approximated to reveal the following relation with the transition rates:
\begin{align}
   P_\pi(t) \approx 
    \left\{\begin{alignedat}{3}
      1 - e^{-\Gamma_\downarrow(t) t_\text{rep}},&\quad &\text{for polarization to $|\text{e}\rangle$},\\
      1 - e^{-\Gamma_\uparrow(t) t_\text{rep}}, &&\text{for polarization to $|\text{g}\rangle$},
    \end{alignedat}\right. \label{eq:Ppi_approx}
\end{align}
with the approximation being of order $\mathcal{O}\left(\Gamma_\downarrow\Gamma_\uparrow t_\text{rep}^2 / 2\right)$, corresponding to the probability that a double quantum jump was undetected within $t_\text{rep}$. We thus obtain $\Gamma_{\uparrow,\downarrow} =  - \log(1 - P_\pi) / t_\text{rep}$, which is the formula used in the main text. We want to remark the following: (i)
when $\Gamma_1$ is constant and known, one can directly solve \eref{eq:Ppi_exact} and obtain both $\Gamma_{\uparrow\downarrow}$-rates, (ii) when $\Gamma_1$ is unknown one can only extract one of the transition rates for each polarization state by solving \eref{eq:Ppi_approx}, and (iii) linearizing both equations gives $\Gamma_{\uparrow,\downarrow} = P_\pi / t_\text{rep}$, however, this approach entails an additional quadratic error compared to solving \eref{eq:Ppi_approx}.

\begin{figure}[H]
\begin{center}
\includegraphics[scale=1]{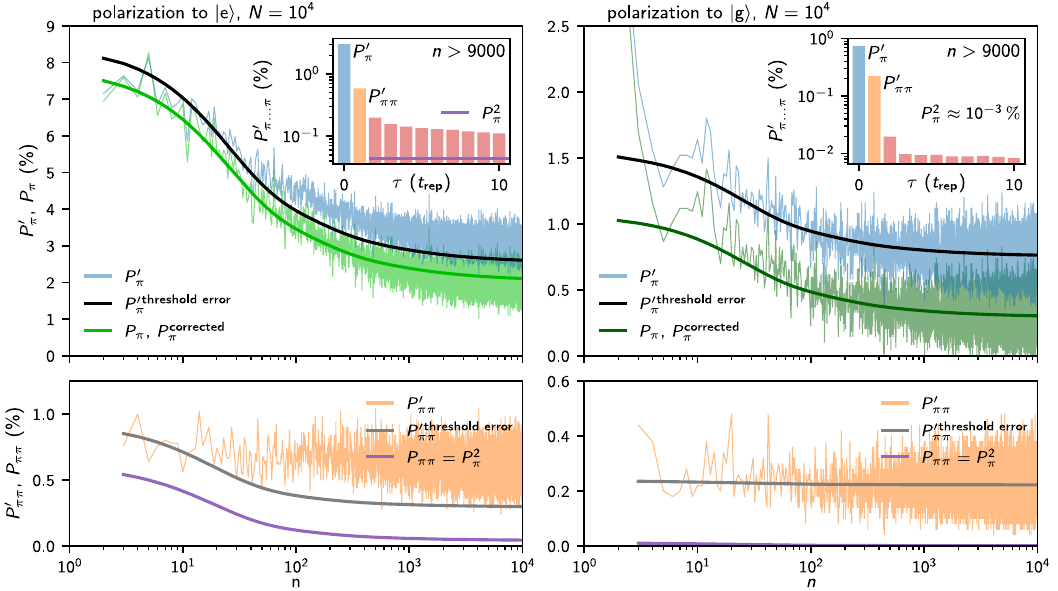}
\caption{\textbf{Active feedback qubit preparations during the polarization sequence.} Measured probabilities $P'_\pi$ (blue) and $P'_{\pi\pi}$ (yellow) during the polarization sequence to play a $\pi$-pulse or respective two successive $\pi$-pulses for polarization to~$|\text{e}\rangle$ (left panels) and polarization to $|\text{g}\rangle$ (right panels). The continuous lines in green and violet show the theoretical expected behavior due to TLS polarization. The theoretical curves in black and grey include state discrimination errors. In light green we show $P_\pi^\text{corrected}$, i.e. the measured $P'_\pi$ corrected for SPAM errors. The insets show the measured auto-correlation function $P'_{\pi\dots\pi}(\tau)$ of the polarization sequence averaged over all $M$ experimental repetitions, over $n > 9000$, where the TLS polarization is approximately constant.
} \label{fig:pi_histogram}
\end{center}
\end{figure}

As mentioned in the introductory paragraph, in the experiment $P_\pi$ is altered by SPAM errors to give the measured $P'_\pi$. In the insets of \fref{fig:pi_histogram} we show the auto-correlation function $P'_{\pi\dots\pi}(\tau) = E[P'_\pi(t + \tau) P'_\pi(t)]$ of the $\pi$-pulse sequence with the expectation value $E$ taken over all $M$ repetitions, over the last one thousand pulses of the sequence where the TLS polarization reached its steady state. In the absence of SPAM errors the $\pi$-pulses are uncorrelated. This means that the probability to have two $\pi$-pulses in succession equals $P'_{\pi\pi} := P'_{\pi\dots\pi}(t_\text{rep}) = {P'_\pi}^2$ and 
more generally $P'_{\pi \text{...}\pi}(\tau > 0) =  {P'_\pi}^2$.
Instead, as we show in \fref{fig:pi_histogram}, we observe a strong excess probability in particular for $P'_{\pi\pi}$. Indeed, if the qubit state was falsely detected or if the reset was unprecise, there will be an increased probability to reset the qubit in the next round. 

We present two approaches to explain and correct SPAM errors. First, we show to which extend the SPAM errors are caused by state discrimination errors, which can easily be included in to the model by forward propagation.
In the experiment the threshold to play a $\pi$-pulse was chosen to be exactly in-between the qubit's pointer states. Therefore, we have a state discrimination error of $p^\text{error}_{|\text{g}\rangle} = 2.42\,\text{\textperthousand}$ and $p^\text{error}_{|\text{e}\rangle} = 2.68\,\text{\textperthousand}$, which was extracted from a Gaussian mixture model fit to the complex scatter parameter $S_{11}$. The difference is due to the slightly squeezed noise as can be seen in \frefadd{fig:high_temp}{c}. (For details about the squeezing see Ref. \cite{Takmakov2022}).
When we include these errors into model \eref{eq:Ppi_exact}, the state discrimination error perfectly explains the polarization to $|\text{g}\rangle$ data, however for polarization to $|\text{e}\rangle$ we only observe a good agreement in the beginning (\fref{fig:pi_histogram}, black and grey curves).
This points to an emerging error as the TLS are increasingly polarized to $|\text{e}\rangle$. 

Since the state discrimination can not account for all SPAM errors, our second approach is to correct the measured $P'_\pi$ by the measured excess probability of $P'_{\pi\pi}$.  Let $P_\pi$ be the probability that the qubit has to be reset due to its relaxation and let $q$ be the probability that the active feedback step yields to a SPAM error. We will assume that an error will be corrected with certainty in the next round, thereby truncating higher order error propagation. We also exclude higher order processes where the qubit relaxed and got falsely measured in its target state. Under these assumptions the probability to find a $\pi$-pulse due to a SPAM error is $P_\text{SPAM} = (1 - P_\text{SPAM}) q$ and hence $P_\text{SPAM} = q / (1 + q)$, which is also the probability to find a $\pi$-pulse that corrects the previous SPAM error. Similarly, the probability to find a $\pi$-pulse due to relaxation is $(1 - P_\text{SPAM}) P_\pi = P_\pi / (1 + q)$. Adding these three probabilities gives $P'_\pi$, while summing the probabilities of all five combinations resulting in two successive $\pi$-pulses gives $P'_{\pi\pi}$. We have:
\begin{align*}
    P'_\pi &= \frac{P_\pi + 2 q}{1 + q} \label{eq_Ppi} \\
    P'_{\pi\pi} &= \frac{P_\pi^2 + 2 q P_\pi + q(1 + q)}{1 + q}.
\end{align*}
Longer ranged correlations are approximately and increasingly more uncorrelated $P'_{\pi\dots\pi}(\tau > t_\text{rep}) \approx {P'_\pi}^2$. The above equations can be inverted allowing to disentangle relaxation and SPAM error contributions:
\begin{alignat*}{3}
    P^\text{corrected}_\pi\,&:=& \;\; P_\pi &= \frac{P'_\pi - (2 - P'_\pi) P'_{\pi\pi}}{(1 - P'_\pi)^2}   \\
    && q &= \frac{P'_{\pi\pi} - {P'_\pi}^2}{(1 - P'_\pi)^2}.
\end{alignat*}
The corrected $P_\pi^\text{corrected}$ curves are shown in \fref{fig:pi_histogram} and match well the theoretical curves $P_\pi$.  These curves are also the ones shown in the main text in \frefadd{fig:1}{c}. At the end of the polarization sequence (from the insets) we extract for the SPAM errors $q = 5.5\,\text{\textperthousand}$ and $q = 2.2\,\text{\textperthousand}$ for polarization in $|\text{e}\rangle$ and $|\text{g}\rangle$, respectively. The accuracy of $P^\text{corrected}_\pi$ and $q$ is limited by all neglected higher order errors and moreover by all the other unexplained excess probabilities starting with $P_{\pi\dots\pi}(2 t_\text{rep})$.

\newpage 
\section{Accounting for state discrimination errors in the extraction of qubit transition rates} \label{app:rates}

\begin{wrapfigure}{r}{8.7cm}
\centering
\includegraphics[scale=1]{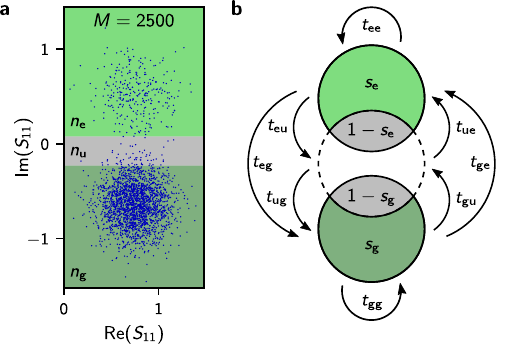}
\caption{\textbf{Extraction of qubit transition rates.} \textbf{a} Typical measured $S_{11}$ scatter plot extracted for a given time index from the $M$ stroboscopic quantum jump traces. The three regions discriminate between areas in the $S_{11}$ plane where we declare the qubit state with certainty larger than \SI{99.98}{\percent}, highlighted in light green for $|\text{g}\rangle$  and dark green for $|\text{e}\rangle$, and where the qubit state is subject to state discrimination errors, highlighted in grey. For measurements in the grey area we declare the state to be undetermined.
\textbf{b} Between two successive sets of $M$ measurements we compute the transition probabilities $t_{ij}$ between the three regions. 
Assuming the qubit is perfectly polarized in $|\text{g}\rangle$ or $|\text{e}\rangle$, the probabilities for measurements to be declared with certainty are given by $s_\text{g}$ and $s_\text{e}$, respectively.
} \label{fig:rates}
\end{wrapfigure}

In the following we detail, how we extract the qubit's population $p_\text{q}$ and transition rates $\Gamma_{\uparrow,\downarrow}$ during the qubit monitoring. The error we have to deal with is the state discrimination error due to the finite signal, meaning that the Gaussian mixtures of the complex reflection coefficient $S_{11}$ overlap (\frefadd{fig:rates}{a}, grey area).

A commonly used approach to extract qubit transition rates from quantum jumps traces consists in analysing the traces sequentially in time, i.e. infer $\Gamma_\uparrow$ and $\Gamma_\downarrow$ by histograming measured durations in-between quantum jumps. However, the time resolution is limited by the time required for a jump to occur. In contrast, here we use a different approach. We acquire a statistically sufficiently large number $M$ of quantum jump traces and we analyze the state distributions at each time index $k$ in the time sequence. 

We measure the qubit stroboscopically with a short (compared to $T_1$) but strong projective readout pulse, therefore with a small probability to measure transitions between pointer states. We define three regions in the complex plane corresponding to the qubit in the ground or excited state, or to an undecided state (\frefadd{fig:rates}{a}).
The regions are chosen in such a way that the probability to mistake the qubit's ground and excited state can be neglected (in our case it may happen in $\approx 0.2\,\text{\textperthousand}$ of the measurements). From the measured $M=2500$ stroboscopic quantum jump traces, we obtain at each time stamp $k$ a distribution similar to the one shown in \frefadd{fig:rates}{a},
and we extract the three population probabilities $p_\text{g}$, $p_\text{e}$ and $p_\text{u}$.
By comparing successive measurement results in each of the $M$ quantum jump traces we extract the nine transition probabilities $t_{ij}$. In other words,
\begin{align*}
    \begin{pmatrix}
         p_\text{g} \\
         p_\text{e} \\ 
         p_\text{u}
    \end{pmatrix}_{k + 1} = 
    \begin{pmatrix}
         t_\text{gg} & t_\text{eg} & t_\text{ug} \\
         t_\text{ge} & t_\text{ee} & t_\text{ue} \\ 
         t_\text{gu} & t_\text{eu} & t_\text{uu}
    \end{pmatrix} \cdot
    {\begin{pmatrix}
         p_\text{g} \\
         p_\text{e} \\ 
         p_\text{u}
    \end{pmatrix}_k}_{\displaystyle \, ,}
\end{align*}
with $\sum_i p_i = 1$ and $\sum_i t_{ij} = 1$. The latter, summing the columns of $t_{ij}$, simply states that the system can either be found in the same region as previously measured or in one of the two other regions. Therefore, out of these three population and nine transition probabilities only eight are independent values. 
Moreover, given the $5.6\,\sigma$ state separation in our measurement, the probability to measure state u is small, the rates $t_\text{ug}$ and $t_\text{ue}$ have low statistical weight and will not be used in the analysis. Nevertheless, we are left with six values, which are sufficient to extract the five parameters of interest: the qubit population $p_\text{q}$, the probabilities $T_{\uparrow\downarrow}$ that the qubit has jumped up or down, and the scaling probabilities $s_\text{g}$ and $s_\text{e}$, which declare the qubit states with certainty.
Instead of computing the most likely set of parameters, there is fortunately a much simpler solution. We have
\begin{align*}
    t_\text{gg} &= (1 - T_\uparrow) s_\text{g} & t_\text{eg} &= T_\downarrow s_\text{g} \\
    t_\text{ge} &= T_\uparrow s_\text{e} & t_\text{ee} &= (1 - T_\downarrow) s_\text{e},
\end{align*}
which can be inverted to give
\begin{align*}
    s_\text{g} &= \frac{t_\text{gg} t_\text{ee} - t_\text{ge} t_\text{eg}}{t_\text{ee} - t_\text{ge}} & s_\text{e} &= \frac{t_\text{gg} t_\text{ee} - t_\text{ge} t_\text{eg}}{t_\text{gg} - t_\text{eg}}.
\end{align*}
Since the scaling probabilities $s_\text{g}$ and $s_\text{e}$ are time independent, they can be averaged along the time trace to achieve a higher precision. 
Finally, we use $s_\text{g}$ and $s_\text{e}$ to obtain the successive quantum jump probabilities corrected for state discrimination errors:
\begin{align*}
    P_{|\text{g}\rangle, |\text{g}\rangle} &= 1 - T_\uparrow = t_\text{gg} / s_\text{g} &  P_{|\text{e}\rangle, |\text{e}\rangle} &= 1 - T_\downarrow = t_\text{ee} / s_\text{e}
\end{align*}
yielding the transition rates (Supp. \ref{app:pi_histogram})
\begin{gather*}
     \Gamma_\uparrow = - \log P_{|\text{g}\rangle, |\text{g}\rangle} / t_\text{rep} \qquad \text{and}  \qquad \Gamma_\downarrow = - \log P_{|\text{e}\rangle, |\text{e}\rangle} / t_\text{rep}.
\end{gather*}
For the qubit population one may either use $p_\text{q} = 1 - p_\text{g} / s_\text{g}$ or $p_\text{q} = p_\text{e} / s_\text{e}$. Ideally, one chooses the regions such that that $s_\text{g} = s_\text{e}$ in which case the most likely estimator for the qubit population is 
\begin{align*}
    p_\text{q} = \frac{n_\text{e}}{n_\text{g} + n_\text{e}}.
\end{align*}

\onecolumngrid

\newpage
\section{Relaxation of the TLS environment} \label{app:tails}

\begin{figure}[H]
\begin{center}
\includegraphics[scale=1.0]{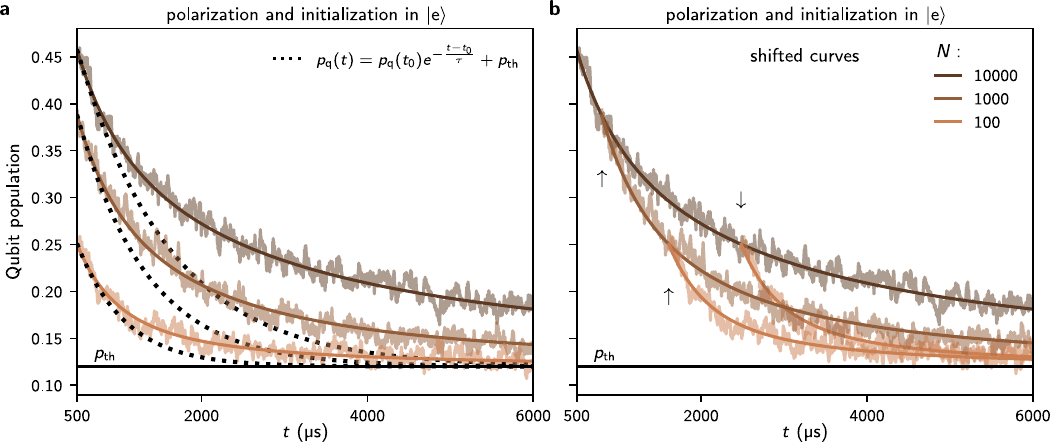}
\caption{\textbf{Relaxation of the environment.} \textbf{a} Relaxation to thermal equilibrium plotted for $t$ starting from \SI{500}{\micro s}, which is an order of magnitude larger than the qubit $T_1$. For comparison, the dotted traces in black show exponential decays which start by matching the measured decay and highlight the slow non-exponential relaxation of the environment at longer time-scales. Notice that we plot the same data as in \frefadd{fig:2}{a} in the main text using a linear time axis. The solid lines show the calculated qubit populations using the same model and fit parameters as in the main text.  \textbf{b} The relaxation tails from panel a shifted in time such that they start at the same qubit population (indicated by the arrow labels). Notice that the initial derivatives differ from each other, highlighting the existence of a memory (hidden variables) in the environment (see discussion in the text).} \label{fig:tails} 
\end{center}
\end{figure}

\begin{figure}[H]
\begin{center}
\includegraphics[scale=1.0]{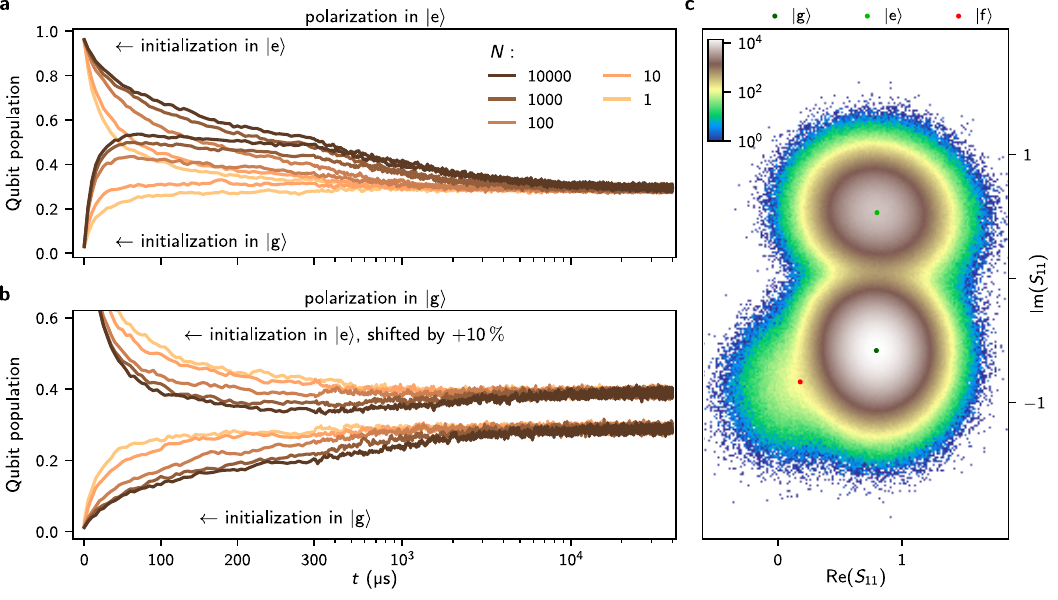}
\caption{\textbf{Relaxation measurements at \textit{T}\;=\;75\,mK}. \textbf{a} Qubit relaxation after polarization to $|\text{e}\rangle$ for various times $N \cdot t_\text{rep}$ followed by an initialization to $|\text{g}\rangle$ or $|\text{e}\rangle$.
\textbf{b} Qubit relaxation after polarization to $|\text{g}\rangle$ followed by an
initialization to $|\text{g}\rangle$ or $|\text{e}\rangle$. Note that at $\SI{75}{mK}$ the final thermal population is higher, therefore increasing the visibility of the cooling effect in comparison to the main text (\frefadd{fig:2}{b}). The curves corresponding to initialization in $|\text{e}\rangle$ are shifted upwards by \SI{10}{\percent} for better visibility. For the stroboscopic qubit monitoring the repetition time was set to $t_\text{rep} = \SI{5}{\micro s}$ in order to decrease the quantum demolition effects (\fref{fig:qnd_effects})
 \textbf{c} Histogram of the complex reflection coefficient $S_\text{11}$ acquired from the last \SI{1}{ms} of the relaxation curves shown in panel a and b. For the thermal populations of the states $|\text{g}\rangle$, $|\text{e}\rangle$ and $|\text{f}\rangle$ we expect $(67.7,~31.9,~\text{and}~0.4)\,\si{\percent}$ and we measure $(70.6\pm0.2,~28.8,~\text{and}~0.4\pm0.2)\,\si{\percent}$. For $|g\rangle$ and $|\text{e}\rangle$ the discrepancy is likely due to quantum demolition effects caused by the readout (\fref{fig:qnd_effects}). The populations are obtained from a Gaussian mixture model fit to the $S_{11}$ distribution using the scikit-learn library~\cite{scikit-learn}.
} \label{fig:high_temp}
\end{center}
\end{figure}

The equilibrium population of the qubit $p_\text{eq}$ is the effective population of its environment. In our case it can be measured in two ways. The first method consists in calculating $p_\text{eq} = \Gamma_\uparrow / \Gamma_1$ with $\Gamma_1=\Gamma_\uparrow+\Gamma_\downarrow$ from the qubit transition rates that can be extracted from quantum jump traces, as described in the main text and shown in \frefadd{fig:3}{c}.
The second method uses the fact that in our case the intrinsic qubit decay is orders of magnitude faster than the relaxation of the TLS environment. Therefore, the tail of the qubit relaxation, i.e. the data shown in \fref{fig:tails}, can directly be ascribed to the effective population of the TLS environment: $p_\text{eq} \approx p_\text{q}$ (see brown and grey curves in \frefadd{fig:3}{c}). The advantage of the latter approach is its superior signal-to-noise ratio.

In \frefadd{fig:tails}{a} we plot the same measured $p_\text{q}$ data as in \frefadd{fig:2}{a} in the main text using a linear time axis instead of the logarithmic axis, in order to highlight the slow non-exponential relaxation.
At this point, one might still imagine that the relaxation curves $p_\text{eq}(t)$ are given by the time evolution of a differential equation $\dot{p}_\text{eq} = f(p_\text{eq})$, where $f$ is not simply proportional to $p_\text{eq}$ (exponential decay) but is an arbitrary function, e.q. similarly to Ref.~\cite{Wang2014}.
In order to rule out this idea, we plot in \frefadd{fig:tails}{b} the relaxation tails from \frefadd{fig:tails}{a} shifted in time such that they start at the same population (indicated by the arrow labels). Clearly, as the derivatives of the relaxation curves differ from each other, the relaxation dynamics can not be described by a first order differential equation of the form $\dot{p}_\text{eq} = f(p_\text{eq})$. Hence, the relaxation must contain hidden variables, i.e. the TLS populations ${p}_\text{t}^k$, that we capture by the system of linear differential equations \eqsref{eq:rate_eq_qubit} and \ref{eq:rate_eq_tls}. 
In general, the equilibrium qubit population $p_\text{eq}$ is defined by $\dot{p_q}=0$ and from \eref{eq:rate_eq_qubit} we have:
\begin{align*}
p_\text{eq}(t) = \frac{\Gamma_\uparrow(t)}{\Gamma_1} = \left( \sum_k  \Gamma_\text{qt}^k p_\text{t}^k(t) + \Gamma_\text{q} p_\text{th} \right) / \, \Gamma_1.
\end{align*}
Consequently, the nonlinear relaxation of the environment originates from the sum over the TLS populations ${p}_\text{t}^k(t)$ that itself are a sum of the $(n + 1)$ exponential solutions of the linear rate equations describing the $n$ TLSs and the qubit. This gives an intuition for the fact that the measured non-exponential relaxation can only be reproduced with a large number of TLSs (more than 15). With an increasing number of TLSs, the agreement improves and the model converges. For all calculations we truncate the model at 101 TLSs, such that $1 / \Gamma_\text{qt}^k$ for $|k| > 50$ is in the order of seconds and much longer than the measured relaxation (Supp.~\ref{app:fitting}).

\newpage
\section{Operating the Szilard engine at \textit{T}\;=\;75\,mK}\label{app:high_temp}
The relaxation curves shown in the main text (\frefadd{fig:1}{a,b}) were measured at a fridge temperature $T = \SI{25}{mK}$. Here, we show measurements at $T = \SI{75}{mK}$ in order to improve the visibility of the cooling effect.

\newpage
\section{Measurements of the TLS environment at different frequencies and in different samples} \label{app:further_TLS_observations}

\begin{figure}[H]
\begin{center}
\includegraphics[scale=1.0]{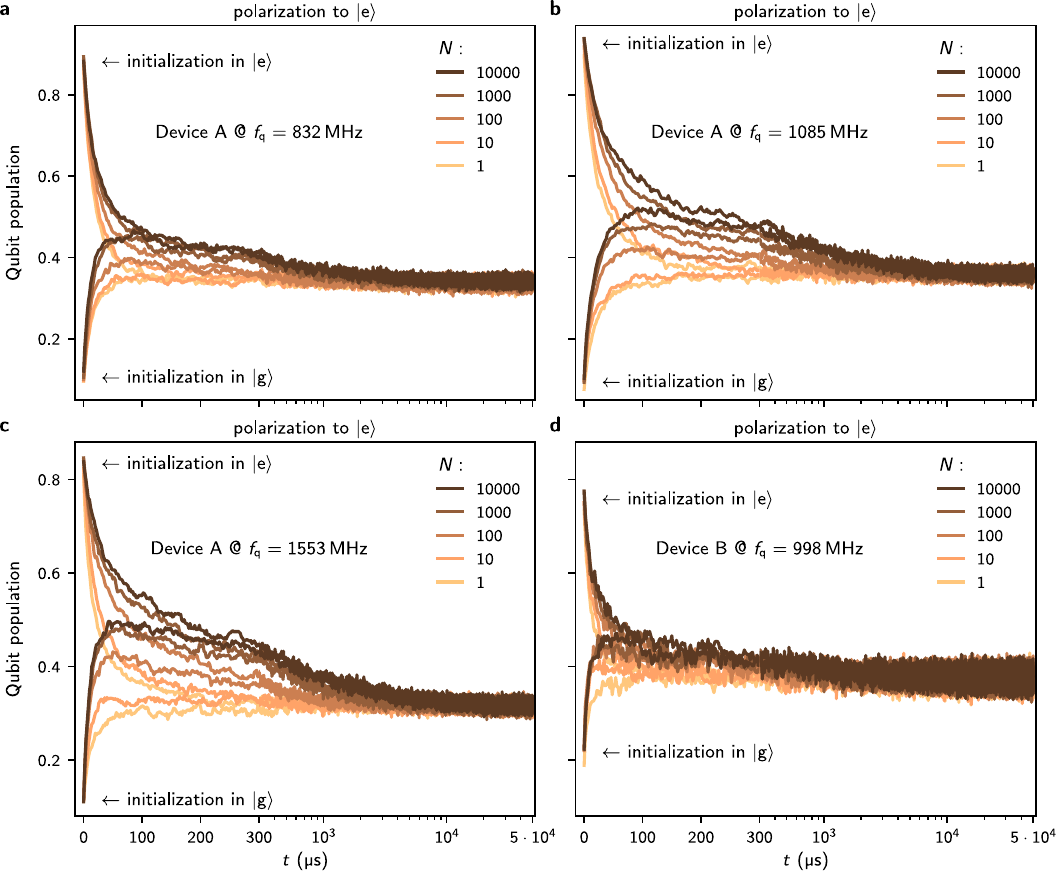}
\caption{\textbf{Observation of the TLS environment at different frequencies in two devices}. Qubit relaxation after polarization to $|\text{e}\rangle$ for various durations, followed by initialization to $|\text{g}\rangle$ or $|\text{e}\rangle$. In panels \textbf{a}, \textbf{b} and \textbf{c} we show measurements on the main text qubit at different frequencies. The flux sweet spot in panel b corresponds to the one in the main text, however notice the qubit frequency has aged.
\textbf{d} Measurements on device B, fabricated on the same wafer as device A.
} \label{fig:further_TLS_observations}
\end{center}
\end{figure}

In \fref{fig:further_TLS_observations} we show additional measurements on the main text sample (device A) at different flux sweet spots corresponding to different qubit frequencies (Supp. \ref{app:fluxonium}). Furthermore, we show in \frefadd{fig:further_TLS_observations}{d} measurements on a second sample (device B), with the same design and fabricated \SI{6}{mm} away from Device A on the same wafer.

The effect of the long-lived TLS environment is visible in all measurements. This supports our model hypothesis that the TLSs are spread in frequency. 
Additionally, comparing the data in \frefadd{fig:further_TLS_observations}{b} to the main text data we see a qualitatively similar relaxation. There are however quantitative differences. In all measurements depicted in \fref{fig:further_TLS_observations} the qubit's idle temperature exceeds $T_\text{eff} > \SI{60}{mK}$. This is due to the use of a new experimental setup and likely caused by the absence of a base plate shield. We note the following main differences: as a consequence of the higher temperature the qubit's $T_1$ decreased to \SI{15.1}{\micro s}, and the qubit's frequency decreased by $\sim \SI{90}{MHz}$ due to junction aging (during $\approx 2.5\,\text{years}$).

\newpage
\section{Details on the fitting procedure} \label{app:fitting}

A priori we have the following independent model parameters: $\Gamma_\text{q}$, $\Gamma_\text{t}$, $a$, $b$, $c$ and the number $n$ of TLSs that we include into the simulation.
In the following we show that $\Gamma_\text{q}$ and $b$ are the most relevant parameters.

We begin our discussion with the number of TLSs. In view of our experimental observation that the TLSs can be found at different frequencies (Supp. \ref{app:further_TLS_observations}), in principle there is no limit on $n$. However, for the simulation we can truncate the far detuned TLSs, which will only contribute at timescales $1 / \Gamma^k_\text{qt}$ on the order of seconds, much longer than the measured relaxation. The model already starts to converge for $\sim 25$ TLSs, however, we increased $n$ to 101 to eliminate any dependence of the slow relaxation tails with the number of TLSs.
The intrinsic relaxation of the TLSs $\Gamma_\text{t} = 1 / 50\,\text{ms}$ is set to its upper bound, which can be deduced from the slow relaxation tail of the $N = 10^4$ curve.
From the measured transition rates we extract the constant qubit relaxation rate $\Gamma_1 = 1/\SI{21.5}{\micro s}$. From the TLS model we have:
\begin{align*}
    \Gamma_1 &= \Gamma_\text{q} + \sum\limits_{k = -\infty}^\infty \frac{a b^2}{b^2 + (k - b\cdot c)^2} \\
    &= \Gamma_\text{q} +  \pi ab \,\frac{\sinh 2 \pi b}{\cosh 2 \pi b - \cos 2 \pi b c}, \label{eq:gamma1_analytic}
\end{align*}
where we replaced the series over $k$ with its closed form. We make use of this relation to express $a$ as as a function of the other fit parameters $\Gamma_\text{q}$, $b$ and $c \in [0, 0.5 / b]$. The advantage of this choice is that we can directly constrain $\Gamma_\text{q} \in [0, \Gamma_1]$ in the fit routine.

\begin{figure}[H]
\begin{center}
\includegraphics[scale=1.0]{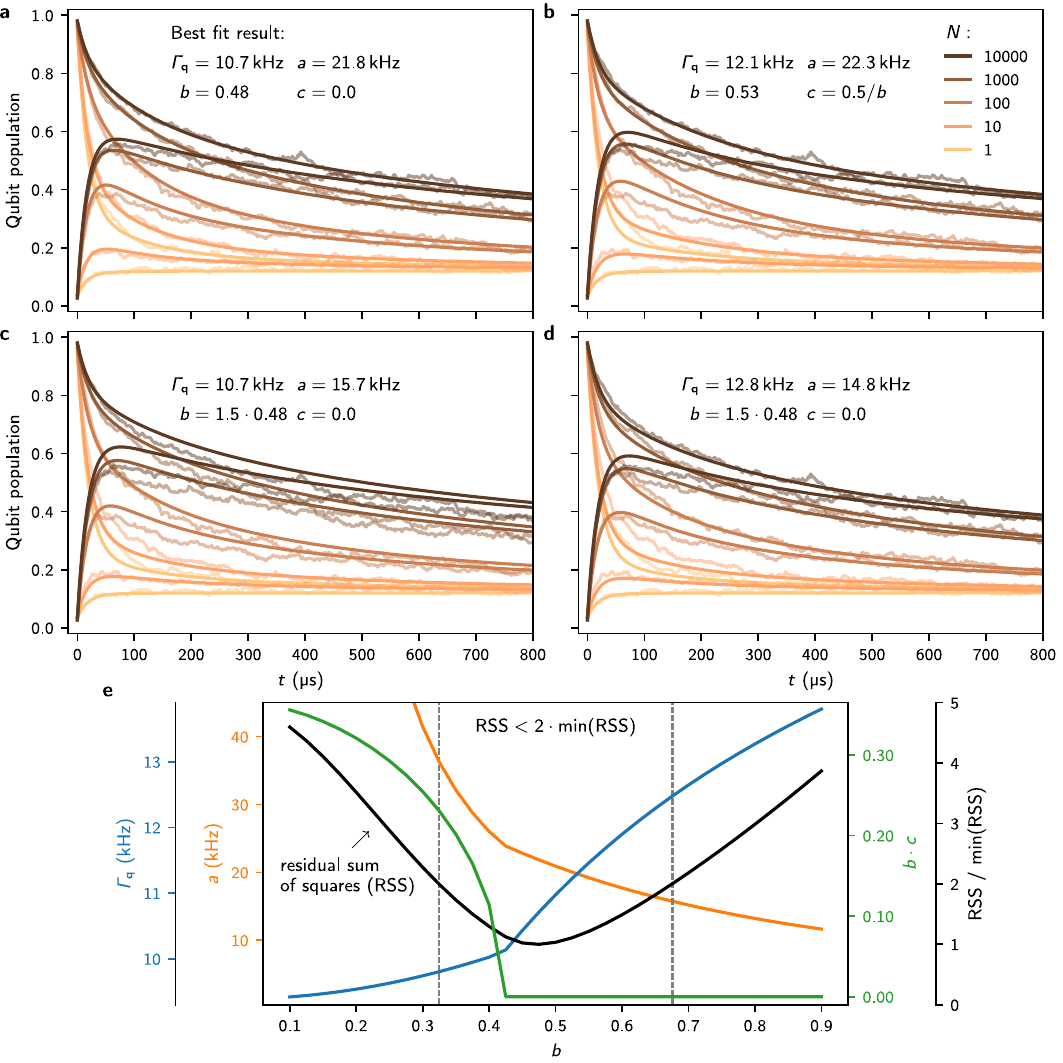}
\caption{\textbf{Influence of the fit parameters on the modeling.}  
The measured data shown in panels a-d is the same as the one presented in \frefadd{fig:2}{a} and the continuous lines show the theoretical model given by \eqsref{eq:gamma_qts}-\ref{eq:rate_eq_tls} with the simplifying assumptions introduced in the main text. \textbf{a} The best fit result, shown for reference. The extracted parameters are the ones stated and used in the main text.
\textbf{b} Influence of the relative frequency shift of the TLS ladder with respect to the qubit frequency. When $c$ is set to its maximum value $c = 0.5 / b$, the agreement with the measurement is worse compared to the fit shown in panel a. As expected, the parameter $c$ changes only the very beginning of the relaxation and
there is only a minor influence on the other fit parameters.
\textbf{c} Influence of the parameter $b$, while keeping $\Gamma_\text{q}$ and $c$ unchanged as in panel a. Note that the parameter $a$ quantifying the TLS qubit coupling has to change by approximately the same ratio as $b$, in order to keep $\Gamma_1$ unchanged. This is expected since $b$ is a measure for the TLS density (in units of $\Gamma_2$).
\textbf{d} Influence of the parameter $b$ on the fit, similar to panel c, but with all fit parameters variable. This allows to fit the slow relaxation dynamics. 
However, the models in panels c and d are inferior in comparison to panel a in two respects. First, the heating of the TLSs takes longer than observed. Second, the cooling effect of the initialization to $|\text{g}\rangle$ is reduced, i.e. the gap between each pair of curves with initialization in $|\text{g}\rangle$ and $|\text{e}\rangle$ is smaller, due to the increased heat capacity of the TLS bath. \textbf{e}
Extracted model parameters and residual sum of squares (RSS) from the fit routine as a function of parameter $b$. The region where $\text{RSS} < 2 \cdot \text{min(RSS)}$ defines the bounds of all fit parameters.
} \label{fig:fit_parameters} 
\end{center}
\end{figure}

In \frefadd{fig:fit_parameters}{a} we show the best fit result (parameters used in the main text). For comparison we show in \frefadd{fig:fit_parameters}{b} that the parameter $c$ only changes the beginning of the relaxation curves, which is expected since the rates of the far detuned TLSs are increasingly unaffected.
The most important parameter is $b$, which encodes the TLS density. If we increase $b$ by 1.5, as shown in \frefadd{fig:fit_parameters}{c} the denser TLSs environment has a larger heat capacitance that has to relax via the qubit as the bottleneck. We can partially compensate this by increasing the qubit relaxation rate, however, as shown in panel \frefadd{fig:fit_parameters}{d} the agreement is not as a good as in panel a. The fact that the model in panel d does not fit as well as the one in panel a is explained by a smaller influence of the qubit initialization on the larger amount of heat stored in TLSs. 

Since the relaxation dynamics occurs on very different time-scales,
the results obtained from the regression depend on the collection of different relaxation curves as well as on the chosen time intervals.
We choose to fit simultaneously all measured relaxation curves shown in \frefadd{fig:fit_parameters}{a} up to \SI{1}{ms} for a given parameter $b$.
The extracted fit parameters as well as the residual sum of squares (RSS) are shown in \frefadd{fig:fit_parameters}{e}. We declare quantitative agreement as long as $\text{RSS} < 2 \cdot \text{min(RSS)}$. The robustness of the model is quantified by the fact that we use only three fit parameters and we obtain quantitative agreement for $\Gamma_\text{q} \in (9.8, 12.5)\,\text{kHz}$, $a \in (15.7, 36.4)\,\text{kHz}$, $b \in (0.32, 0.68)$, $b \cdot c \in (0, 0.23)$ and $\Gamma_\text{t} \in (0, 20)\,\text{Hz}$.
Finally, we emphasize that even though the fit parameters are fixed by the first one millisecond, the model describes the entire non-exponential relaxation up to 50ms (Supp. \ref{app:tails}), cementing its validity.

\newpage
\section{Coherence properties of the qubit} \label{app:coherence}
\begin{figure}[H]
\begin{center}
\includegraphics[scale=1]{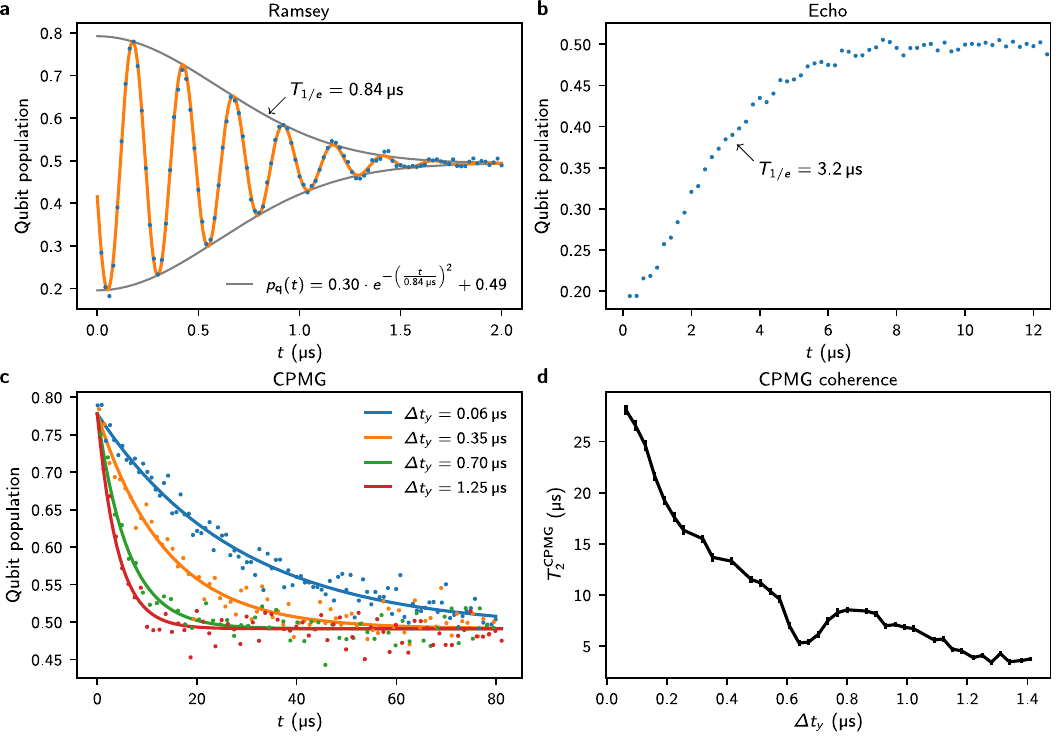}
\caption{\textbf{Coherence measurements of the qubit.} \textbf{a} Measured Ramsey decay with Gaussian envelope (blue points are measured and solid lines are fits). \textbf{b} Echo decay, which contains both exponential and Gaussian components. The arrow indicates the $1/e$-time. \textbf{c} Selection of measured CPMG decays (points) with exponential fits (solid lines). \textbf{d} Coherence times as a function of the $y$-pulse repetition time in the CPMG sequence. The coherence times and the corresponding standard deviation errors are obtained from the exponential fits (panel c). The lines connecting the markers with error bars are guides to the eye.
} \label{fig:coherence}
\end{center}
\end{figure}

In \fref{fig:coherence} we show various coherence measurements. While the Gaussian envelope for the Ramsey sequence in \frefadd{fig:coherence}{a} reveals a significant contribution of slow noise, the coherence increases significantly with the number of refocussing $\pi$-pulses (\frefadd{fig:coherence}{b,c}). The decay of the Carr-Purcell-Meiboom-Gill (CPMG) sequence is well described by an exponential function with a maximum coherence exceeding $T_1 = \SI{21.5}{\micro s}$ (\frefadd{fig:coherence}{d}). Since the coherence is closely related to the qubit's frequency noise, the dip visible at $f = 1 / 2 \Delta t_y = \SI{0.77}{MHz}$ indicates an excess noise source. 
While we believe that most of the frequency noise comes from flux noise, it is possible that the TLSs also contribute and this hypothesis should be addressed in future studies.

\newpage
\section{The quantum Szilard engine} \label{app:szilard}

\begin{figure}[H]
\begin{center}
\includegraphics[scale=1.0]{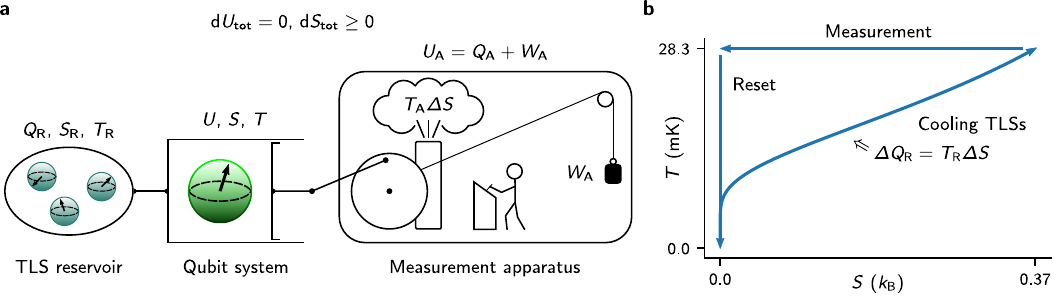}
\caption{\textbf{The quantum Szilard engine.} \textbf{a} Sketch of the whole thermodynamic system consisting of the measurement apparatus, the qubit and the TLS reservoir. For a closed system the internal energy is constant while the entropy can only increase irreversibly according to the second law of thermodynamics. The measurement and the feedback is performed by the intelligent being, as conceived by Szilard~\cite{Szilard1929}. \textbf{b} Refrigeration cycle of the Szilard engine cooling the TLS reservoir.} \label{fig:szilardengine} 
\end{center}
\end{figure}

\begin{figure}[H]
\begin{center}
\includegraphics[scale=1.0]{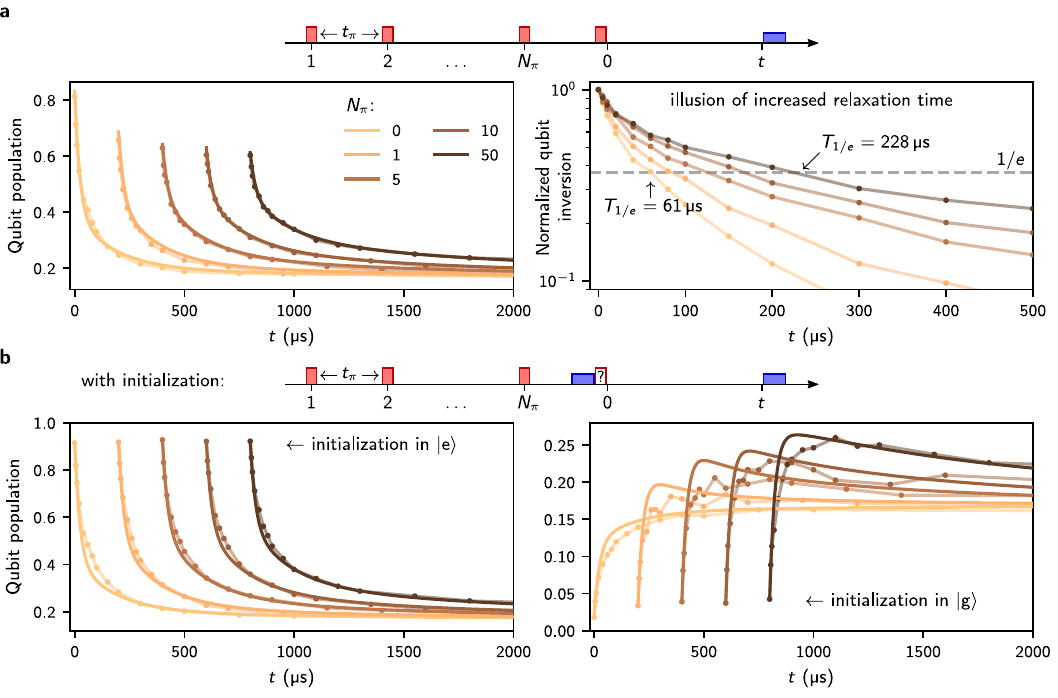}
\caption{\textbf{ Heating without active feedback.} \textbf{a} Top panel: schematic of the pulse sequence consisting of $N_\pi$ repeated $\pi$-pulses (red boxes) spaced by $t_\pi = \SI{100}{\micro s}$, followed by a free decay measurement (the measurement pulse is indicated by the blue box). The value for $t_\pi$ is chosen larger than the intrinsic qubit decay time but smaller than the relaxation of the environment, in order to heat the environment.
Left panel: Free decay of the qubit for various $N_\pi$. The curves are shifted horizontally for visibility. Right panel: Measured relaxation curves taken from the left panel, normalized and plotted in log-scale. For our device, the ostensibly increased relaxation time is an illusion, and it is explained by the increased environmental TLS population which heats the qubit, as demonstrated in panel b (right panel) and in the main text \fref{fig:3}. Consequently, as also discussed in Supp. \ref{app:tails}, this heating of the environment forbids us to compare scaled and shifted non-exponential relaxation curves. \textbf{b} Top panel: schematic of the pulse sequence consisting of $N_\pi$ repeated $\pi$-pulses spaced by $t_\pi = \SI{100}{\micro s}$, followed by an initialization to $|\text{g} \rangle$ or $|\text{e}\rangle$ immediately before the free decay measurement. Notice that this sequence is identical to the one in panel a, with the exception of the initialization pulse.
The left panel measurements after initialization in $|\text{e}\rangle$ appear similar to the corresponding ones in panel a while,
strikingly, after initialization to $|\text{g}\rangle$ (right panel) we observe non-monotonic evolutions of the qubit population, due to the heat stored in the TLS environment. The solid lines are simultaneous fits using the theoretical model of \eqsref{eq:rate_eq_qubit} and \ref{eq:rate_eq_tls}, including the $\pi$-pulse sequence on the qubit.
} \label{fig:pi_pulse_heating} 
\end{center}
\end{figure}

For sake of completeness, we briefly discuss the thermodynamic properties of the Szilard engine focusing on its usage as a refrigerator. The whole thermodynamic system and the refrigeration cycle are depicted in \fref{fig:szilardengine}.
Following Szilard~\cite{Szilard1929}, we consider a system consisting of a ground state and an excited state with a $d$-fold degeneracy. For $d = 1$ we have the experimental situation where the system can be referred to as the qubit. The system's internal energy $U$ and entropy $S$ read:
\begin{align*}
U = \frac{d\epsilon}{d + e^{\beta \epsilon}} \quad \text{and} \quad
S / k_\text{B} = (S_\text{rev} + S_\text{irr}) / k_\text{B}  = \beta U + \ln\left(d + e^{\beta \epsilon}\right) - \beta \epsilon,
\end{align*}
where $\epsilon$ is the energy spacing between the two system states, $\beta = 1 / k_\text{B} T$ encodes the temperature $T$ of the system  and $k_\text{B}$ is the Boltzmann constant. The entropy can be divided into two components. The reversible entropy $S_\text{rev}$ can be exchanged with the TLS reservoir, while the irreversible entropy $S_\text{irr}$ can only increase during a thermodynamic process and is closely related to the free energy $F = T S_\text{irr}$ of the system.\\

When the system is measured quantum mechanically with an operator that collapses the system state either to its ground or excited state manifold, the entropy reduces, depending on the measurement outcome, to $S /k_\text{B}  = 0 $ or $S /k_\text{B} = \ln d$, respectively. The maximum average entropy reduction of $\Delta S /k_\text{B} = \ln 2$ is attained when $\beta \epsilon = \ln d$. For the qubit, where $d = 1$, this value is only reached in the limit of an infinite temperature or a vanishing energy level splitting~\cite{Pekola2016}.

As we consider the measurement apparatus to be a thermodynamic engine, the entropy reduction has to be compensated so that the second law of thermodynamics remains valid. Consequently, the apparatus must be connected to a heat bath to which it can unload at least the reduced entropy. When this bath is at the temperature $T_\text{A}$, the measurement requires the minimum work $W_\text{M} = T_\text{A} \Delta S$. Furthermore, one can argue that the measurement process should not depend on the temperature of the system. Thus, $W_\text{M} \geq k_\text{B} T_\text{A}\ln 2$ must hold, as was first conjectured by Szilard.
For the performance consideration of the refrigerator we will drop this assumption and use the exact entropy reduction to allow for a simple comparison with the theoretical maximum performance given by the Carnot cycle.

After the measurement, the information on the system state can be used to cool down the TLS reservoir in which case the system has to be reset to its ground state.
Here, we need an additional discussion for systems with a degenerate excited state. When the system is measured in the excited state manifold it cannot simply be reset to its ground state as it would mean to destroy the remaining entropy $S /k_\text{B} = \ln d$. Instead, additional measurements are required to determine the exact state of the system allowing to select the correct gate operations bringing the system to its ground state. Alternatively, one could think of a more powerful measurement that can distinguish between all $(d + 1)$-states. This measurement, however, can produce a maximum average entropy reduction of $\Delta S /k_\text{B} = \ln (d + 1)$ in the limit $\beta \epsilon \rightarrow 0$, in accordance with the previously mentioned limit for the qubit.

Despite, these technical details concerning the reset of the degenerate excited state to its ground state, the whole thermodynamic cycle can be summarized in three steps:
\begin{enumerate}
    \item The measurement requires the work $W_\text{M} = T_\text{A} \Delta S$.
    \item  From the reset of the qubit one can in principle extract the work $W_\text{Q} = - \Delta U$.
    \item  The reservoir is cooled by the amount $\Delta Q_\text{R} = T_\text{R} \Delta S_\text{rev} = \Delta U$. Here, we assume that the TLS reservoir is large enough so that its temperature $T_\text{R}$ stays approximately constant.
\end{enumerate}
The coefficient of performance (COP) now reads:
\begin{align*}
    \text{COP} = \frac{\Delta Q_\text{R}}{W_\text{tot}} = \frac{T_\text{R} \Delta S_\text{rev}}{T_\text{A} (\Delta S_\text{rev} + \Delta S_\text{irr}) - T_\text{R} \Delta S_\text{rev}} = \frac{T_\text{R}}{T_\text{A} - T_\text{R}  + T_\text{A} \frac{\Delta S_\text{irr}}{\Delta S_\text{rev}}}, 
\end{align*}
showing that the Szilard engine will always operate below the maximum theoretical efficiency, which is only reached in the following limit:
\begin{align*}
\frac{\Delta S_\text{irr}}{\Delta S_\text{rev}} = \frac{d + e^{\beta \epsilon}}{d} \left(\frac{\ln\left(d + e^{\beta \epsilon}\right)}{\beta \epsilon } - 1\right) = 0 +  \mathcal{O}\left(\frac{1}{\beta\epsilon}\right) \quad \text{for} \quad \beta\epsilon\rightarrow\infty,
\end{align*}
while in contrast the cooling power given by $\Delta U$ vanishes exponentially.\\

In our experiment, the reservoir can be cooled at most by $\Delta Q_\text{R} \approx 0.5 \Delta U$ as stated in the main text. This surprisingly small value seems to be in conflict with the TLS bath being the dominant loss mechanism $\Gamma_\text{qt}^\text{TLSs} \approx 3.3 \Gamma_\text{q}$. This discrepancy can simply be explained by the finite size of the reservoir. The qubit only interacts strongly with the few most resonant TLSs. Consequently, when the qubit is reset to its ground state the temperature of these TLSs will reduce and the qubit can not reach its prior energy, thus $\Delta Q_\text{R} < \Delta U(T_\text{R})$.

\newpage
\section{Heating without active feedback} \label{app:pi_pulse_heating}

Here, we show that the reservoir can be heated by a sequence of $\pi$-pulses. Our results resemble those reported in Ref.~\cite{Gustavsson2016Dec}, however, at least for our qubit the seemingly increased $T_1$-time is simply due to the heated environment. Despite the similarities, the environments probed in the two experiments are not necessarily of the same nature.

\end{document}